\documentclass[aps,twocolumn]{revtex4}
\usepackage{bm}
\usepackage{amsmath}
\usepackage{amsmath,amssymb}
\allowdisplaybreaks
\usepackage{graphicx}
\usepackage{multirow}
\newcommand{\dis}{\displaystyle}
\newcommand{\mean}[1]{\langle{#1}\rangle}
\newcommand{\bra}[1]{\langle{#1}|}
\newcommand{\ket}[1]{|{#1}\rangle}

\begin{document}
\title{Quantum functionalities via feedback amplification}
\author{Rion Shimazu}
\email[]{shark98t@keio.jp}
\author{Naoki Yamamoto}
\email[]{yamamoto@appi.keio.ac.jp}
\affiliation{Department of Applied Physics and Physico-Informatics, 
Keio University, Hiyoshi 3-14-1, Kohoku, Yokohama 223-8522, Japan}
\date{\today}

\begin{abstract}
Feedback amplification is a key technique for synthesizing various important 
functionalities, especially in electronic circuits involving op-amps. 
This paper presents a quantum version of this methodology, where the general 
phase-preserving quantum amplifier and coherent (i.e., measurement-free) feedback 
are employed to construct various type of systems having useful functionalities: 
quantum versions of differentiator, integrator, self-oscillator, and active filters. 
The class of active filters includes the Butterworth filter, which can be used to 
enhance the capacity of an optical quantum communication channel, and 
the non-reciprocal amplifier, which enables measurement of a superconducting qubits system 
as well as protection of it by separating input from output fields. 
A particularly detailed investigation is performed on the active phase-cancelling filter 
for realizing a broadband gravitational-wave detector; 
that is, the feedback amplification method is used to construct an active filter that 
compensates the phase delay of the signal and eventually recovers the sensitivity 
in the high frequency regime. 
\end{abstract}

\maketitle


\section{Introduction}
\label{sec:Introduction}

\begin{figure}[t]
\centering 
\includegraphics[width=8.4cm]{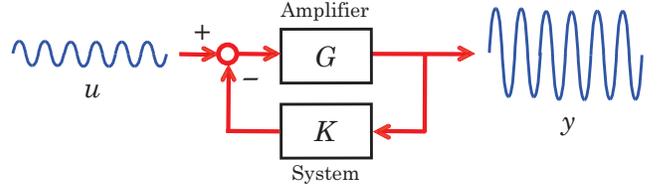}
\caption{
Schematic of the classical feedback amplifier. 
}
\label{classical fb}
\end{figure} 

The amplifier is an essential component in modern technological systems, and it is 
usually involved in those systems in some feedback form. 
Let us consider a classical amplification process $y=Gu$ where $u$ and 
$y$ are input and output signals, and $G>1$ is the gain of the amplifier. 
Then by feeding a fraction of the output back to the input through the controller 
$K$, as depicted in Fig.~\ref{classical fb}, the input-output relation is 
modified to 
\begin{equation}
     y=G^{({\rm fb})} u,  ~~~ 
     G^{({\rm fb})}=\frac{G}{1+GK}=\frac{1}{1/G + K}.\notag
\end{equation}
Then by making the gain $G$ large, we find $y=(1/K)u$; hence if $K$ is a passive 
device with gain $K<1$, the entire system works as a robust amplifier which is 
insensitive to the parameter change in $G$. 
The importance of this feedback amplification technique \cite{Black 77, Black 84} 
is not limited to realizing of such a robust amplifier. 
That is, by combining high-gain amplifiers (op-amps in the electrical circuits) 
with several passive devices such as resistors and capacitors, one can devise a 
variety of functional systems; e.g., integrator, active filters, switches, and self-oscillators 
\cite{op-amp textbook}.

This paper develops the quantum version of feedback amplification theory, which is 
expected to be of particular importance to make the existing quantum technological 
devices robust and further to engineer systems with functionalities. 
In fact this idea has been implicitly employed in some specific systems 
\cite{Courty 99, Clerk 2010}. An explicit research direction was addressed in 
\cite{Yamamoto 2016}, showing 
a general quantum analogue to the above-described robust amplification method; 
more precisely, it is shown that a coherent (i.e., measurement-free) feedback 
control \cite{Wiseman 94,Yanagisawa 2003,Mabuchi 2008,James 2008,Gough 09,
Mabuchi 2012,Kerckhoff 2013,Yamamoto 14} of a high-gain phase-preserving amplifier 
\cite{Haus 62,Caves 1982,Clerk 2010,Devoret 10,Caves 2012} and a passive device 
(e.g., a beam splitter) yields a robust phase-preserving amplifier.

This paper begins with Sec.~II to introduce the models of the quantum phase-preserving amplifier 
and some linear passive systems. Then, using those models, 
we extend the quantum feedback amplification scheme presented in \cite{Yamamoto 2016} from 
the Fourier domain to the Laplace domain (Sec.~III), together with developing a basic stability test 
method (Sec.~IV). 
We then apply the theory to construct systems having several useful functionalities: 
quantum versions of differentiator and integrator (Sec.~V), self-oscillator 
(Sec.~VI), and active filters (Sec.~VII). 
As for the quantum integrator, it will be proven applicable for improving the 
detection efficiency of an itinerant field. 
The ability to synthesize a quantum self-oscillator might also be useful for several purposes 
as in the classical case, such as analogue quantum memory and frequency converter 
\cite{Mabuchi 15, Safavi-Naeini 19}, though in this paper we do not provide a concrete example. 
Active filtering is a typical application of feedback amplification, which in our 
case includes the quantum version of Butterworth filter \cite{Laghari 2014} and 
non-reciprocal amplifier; 
the former is used to realize the steep roll-off characteristic in frequency, 
which enables the enhancement of the capacity of a quantum communication channel 
\cite{Shapiro 2016}; the latter enables precise measurement of a superconducting qubits system 
while protecting it from the unwanted backward field generated in the amplification process 
\cite{Yurke, Abdo 14, Clerk 15,Malz 2018,Abdo 2018}.

In particular, in Sec.~VIII we show a detailed investigation on the quantum phase-cancellation 
filter applied to the gravitational-wave detection problem; 
this is an active filter that can compensate the delayed phase of an incoming signal 
for the purpose of enhancing the detection bandwidth. 
The quantum phase-cancellation filters proposed in the literature 
\cite{Miao 2015,Shahriar 2018,Blair 2018,Miao 2019} are based on an opto-mechanical 
implementation, but it requires an extremely low environmental temperature. 
The proposed phase-cancellation filter based on the feedback amplification method, on the 
other hand, can be all-optically implemented in the room temperature. 
We demonstrate a numerical simulation to show how much this filter can broaden the 
bandwidth of a typical gravitational-wave detector in a practical setting.


\section{Preliminaries}
\label{sec:Preliminaries}

\subsection{Phase preserving linear amplifier}
\label{sec:Phase preserving linear amplifier}

\begin{figure}[t]
\centering\includegraphics[width=0.6\columnwidth]{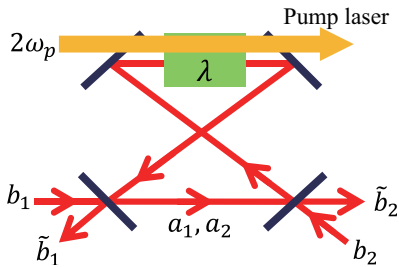}
\caption{Non-degenerate parametric amplifier}
\label{NDPA}
\end{figure}

In this paper we consider a general phase preserving linear amplifier 
\cite{Haus 62,Caves 1982, Caves 2012}. 
A typical realization of this system is given by the {\it non-degenerate parametric 
amplifier (NDPA)} \cite{Clerk 2010, Ou}. 
In optics case, as depicted in Fig.~\ref{NDPA}, the NDPA is an optical cavity having 
two orthogonally polarized fields with modes $a_1$ and $a_2$, which are created and 
coupled with each other at the pumped non-linear crystal (the green box in 
Fig.~\ref{NDPA}) inside the cavity. 
Also, the mode $a_1$ ($a_2$) couples with an input field $b_1$ ($b_2$) at the mirror 
with transmissibity proportional to $\gamma$. 
The Hamiltonian of the NDPA is given by
\begin{align}
    H_\textrm{NDPA}=&\hbar\omega_1a_1^\dag a_1+\hbar\omega_2a_2^\dag a_2\notag\\
    &+i\hbar\lambda (a_1^\dagger a_2^\dagger e^{-2i\omega_pt} - a_1a_2e^{2i\omega_pt}),\notag
\end{align}
with $\omega_k$ the resonant frequencies of $a_k$, $\lambda\in\mathbb{R}$ the coupling 
strength between $a_1$ and $a_2$, and $2\omega_p$ the pump frequency. 
Here we assume that $\omega_1=\omega_2=\omega_p$. 
Then, in the rotating frame at frequency $\omega_p$, the dynamics of the NDPA is given 
by the following Langevin equation \cite{Gardiner Book}:
\begin{equation}
            \left[ \begin{array}{c}
                \dot{a}_1 \\ 
                \dot{a}_2^\dagger \\
            \end{array} \right]
       = \left[ \begin{array}{cc}
                -\gamma/2 & \lambda \\ 
                \lambda & -\gamma/2 \\
            \end{array} \right]
            \left[ \begin{array}{c}
                a_1 \\ 
                a_2^\dagger \\
            \end{array} \right]
           -\sqrt{\gamma}
            \left[ \begin{array}{c}
                b_1 \\ 
                b_2^\dagger  \\
            \end{array} \right]. 
\label{dynamics of NDPA}
\end{equation}
Note that the canonical commutation relation of input fields is given by 
$[b(t), b^\dag(t')]=\delta(t-t')$, with $\delta(t-t')$ the Dirac delta function. 
The output equations are given by
\begin{equation}
     \tilde{b}_1=\sqrt{\gamma}a_1 + b_1,~~~
     \tilde{b}_2^\dag=\sqrt{\gamma}a_2^\dag + b_2^\dag.
\label{output equations of NDPA}
\end{equation}
From Eqs. \eqref{dynamics of NDPA} and \eqref{output equations of NDPA}, 
the input-output relation of the NDPA is represented as 
\begin{eqnarray}
\label{transfer function of NDPA}
& & \hspace*{-1em}
      \left[ \begin{array}{c}
                \tilde{b}_1(s) \\ 
                \tilde{b}_2^\dagger(s^\ast) \\
            \end{array} \right]
       = \frac{1}{(s+\gamma/2)^2-\lambda^2}
\nonumber \\ & & \hspace*{0em}
        \times\left[ \begin{array}{cc}
                s^2 -\lambda^2 - \gamma^2/4 & -\gamma \lambda \\
                -\gamma \lambda & s^2 -\lambda^2 - \gamma^2/4 \\
            \end{array} \right]            
            \left[ \begin{array}{c}
                b_1(s) \\ 
                b_2^\dagger(s^\ast) \\
            \end{array} \right]. 
\end{eqnarray}
The operator $b(s)$ is related to $b(t)$ via the Laplace transformation 
\cite{Yamamoto 2016,Nurdin NY book,Gough 2017 arxiv}: 
\[
	b(s)=\int_0^\infty e^{-st}b(t)dt,\quad 
        b^\dag(s)=[b(s)]^\dag=\int_0^\infty e^{-s^\ast t}b^\dag(t)dt.
\]
From Eq.~\eqref{transfer function of NDPA}, $\gamma>2\lambda$ if and only if 
the amplifier is stable (i.e., every solution of the characteristic polynomial 
$(s+\gamma/2)^2-\lambda^2=0$ has negative real part). 
The output mode $\tilde{b}_1$ at $s=0$ is given by 
\[
     \tilde{b}_1(0) 
        = -\frac{\gamma^2 + 4\lambda^2}{\gamma^2 - 4\lambda^2} b_1(0) 
          + \frac{-4\gamma \lambda}{\gamma^2 - 4\lambda^2} b_2^\dagger(0), 
\]
which diverges as $\gamma\rightarrow 2\lambda+0$. 
Hence, in this parameter limit, the signal with $s$ satisfying $|s|\ll \gamma$ 
is largely amplified.

In this paper we consider the general phase-preserving linear amplifier with 
the following input-output relation: 
\begin{align}
      \left[ \begin{array}{c}
                \tilde{b}_1(s) \\ 
                \tilde{b}_2^\dagger(s^\ast) \\
            \end{array} \right]
        =& ~ G(s)
            \left[ \begin{array}{c}
                b_1(s) \\ 
                b_2^\dagger(s^\ast) \\
            \end{array} \right],\notag\\
        G(s)=&\left[ \begin{array}{cc}
             G_{11}(s) & G_{12}(s) \\
             G_{21}(s) & G_{22}(s) \\
            \end{array} \right].
\label{input output relation of NDPA}
\end{align}
The condition on the transfer function matrix $G(s)$ is represented in the 
Fourier domain as follows. 
The Fourier transformation of the field operators are defined as
\begin{align}
	&b(i\omega)=\int_{-\infty}^\infty e^{-i\omega t}b(t)dt,\notag\\ 
        &b^\dag(i\omega)=[b(i\omega)]^\dag
            =\int_{-\infty}^\infty e^{i\omega t}b^\dag(t)dt, \notag
\end{align}
which satisfy $[b(i\omega),b^\dag(i\omega')]=2\pi\delta(\omega-\omega')$. 
This commutation relation requires $G(s)$ to satisfy 
\begin{align}
    &|G_{11}(i\omega)|^2 - |G_{12}(i\omega)|^2 
    = 
       |G_{22}(i\omega)|^2 -|G_{21}(i\omega)|^2 = 1, \notag
 \\
   & G_{21}(i\omega)G_{11}^*(i\omega) - G_{22}(i\omega)G_{12}^*(i\omega)=0,  
    \quad\forall\omega, 
\label{G condition}
\end{align}
where $G_{ij}^*(i\omega)=[G_{ij}(i\omega)]^\ast$ is the complex conjugate of 
$G_{ij}(i\omega)$.


\subsection{Passive systems}
\label{sec:Passive components}

\begin{figure}[t]
\centering
\leftline{\hspace{2em}\normalsize(a)\hspace{10.5em}\normalsize(b)}
\includegraphics[width=7.5cm]{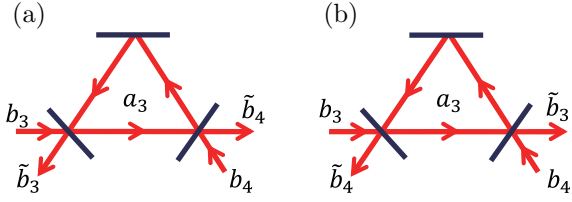}
\caption{
Single-mode optical cavity, functioning as (a) the low-pass filter and (b) the high-pass filter, 
for the input-output relation from $b_3$ to $\tilde{b}_4$. }
\label{MCCs}
\end{figure} 

The general form of passive linear system from the inputs $(b_3, b_4)$ to 
the outputs $(\tilde{b}_3, \tilde{b}_4)$ in the Laplace domain is represented as 
\begin{align}
      \left[ \begin{array}{c}
                \tilde{b}_3^\dagger(s^\ast) \\ 
                \tilde{b}_4^\dagger(s^\ast) \\
            \end{array} \right]
        =&K(s) 
            \left[ \begin{array}{c}
                b_3^\dagger(s^\ast) \\ 
                b_4^\dagger(s^\ast) \\
            \end{array} \right],\notag\\
        K(s)=& \left[ \begin{array}{cc}
             K_{11}(s) & K_{12}(s) \\
             K_{21}(s) & K_{22}(s) \\
            \end{array} \right],
            \label{input output relation of MCC}
\end{align}
where the creation-mode representation is used to simplify the notation. 
The transfer function $K(s)$ satisfies $|K_{11}(i\omega)|^2 + |K_{12}(i\omega)|^2 = 1$, 
$|K_{21}(i\omega)|^2 + |K_{22}(i\omega)|^2 = 1$, and 
$K_{21}(i\omega)K_{11}^*(i\omega) + K_{22}(i\omega)K_{12}^*(i\omega)=0$, 
$\forall \omega$. 
These conditions are derived from unitarity of the response function matrix of 
passive quantum system:
\begin{equation}
    K^{-1}(i\omega) = K^\dag (i\omega) =
         \left[ \begin{array}{cc}
             K_{11}^\ast (i\omega) & K_{21}^\ast (i\omega) \\
             K_{12}^\ast (i\omega) & K_{22}^\ast (i\omega) \\
            \end{array} \right]. \notag \label{eq: unitary passive systems}
\end{equation}

A typical passive device is a single-mode optical cavity having two input-output 
ports, depicted in Fig.~\ref{MCCs}(a). 
The dynamics of the cavity is given by 
\begin{equation}
    \dot{a}_3^\dag =  \Big(- \frac{\kappa_1+\kappa_2}{2} + i\Delta\Big)a_3^\dag
    - \sqrt{\kappa_1}b_3^\dag - \sqrt{\kappa_2}b_4^\dag,\notag
\end{equation}
where $a_3$ is the cavity mode, $\kappa_i$ is the coupling strength between $a_3$ and 
the input itinerant field $b_i$, and $\Delta$ is the detuning. 
Also the output equations are given by 
\[
     \tilde{b}_3^\dag=\sqrt{\kappa_1}a_3^\dag + b_3^\dag,~~~
     \tilde{b}_4^\dag=\sqrt{\kappa_2}a_3^\dag + b_4^\dag.
\]
Then the transfer function matrix $K(s)$ is given by
\begin{align}
       K&(s) =  \frac{1}{s+(\kappa_1+\kappa_2)/2-i\Delta}\notag \\ 
       &\times    
         \left[ \begin{array}{cc}
                s + (\kappa_2 - \kappa_1)/2 -i\Delta & -\sqrt{\kappa_1\kappa_2} \\
                -\sqrt{\kappa_1\kappa_2} & s + (\kappa_1 - \kappa_2)/2 -i\Delta \\
            \end{array} \right].
            \label{asymmetric LPF}
\end{align}
In the special case $\kappa_1=\kappa_2=\kappa$ and $\Delta=0$, it is 
\begin{equation}
\label{LPF}
       K(s)       
        =  \frac{1}{s+\kappa}
            \left[ \begin{array}{cc}
                s  & -\kappa \\
                -\kappa & s  \\
            \end{array} \right].
\end{equation}
Hence the relation between $b_3$ and $\tilde{b}_4$ is given by 
\begin{equation}
\label{passive cavity input output relation 1}
     \tilde{b}_4^\dag(s^\ast) 
         = \frac{-\kappa}{s+\kappa} b_3^\dag(s^\ast)
         + \frac{s }{s+\kappa} b_4^\dag(s^\ast).
\end{equation}
That is, in the domain $|s|\gg \kappa$, the cavity works as an integrator for 
the transmitting field from $b_3$ to $\tilde{b}_4$. 
Also it works as a low-pass filter with bandwidth $\kappa$; that is, the frequency 
components of $b_3$ with $|s|=|i\omega|\ll\kappa$ can only pass through the cavity, 
and hence this cavity is called the mode-cleaning cavity (MCC). 

In this paper we also work on the case where $\tilde{b}_4$ is the reflected field of $b_3$, 
as shown in Fig.~\ref{MCCs}(b); 
in this case the transfer function is given by 
\begin{eqnarray}
\label{asymmetric HPF}
& & \hspace*{-0.9em}
       K(s) =  \frac{1}{s+(\kappa_1+\kappa_2)/2-i\Delta}  \nonumber \\ 
& & \hspace*{1.2em}
        \times 
            \left[ \begin{array}{cc}
                -\sqrt{\kappa_1\kappa_2} & s + (\kappa_1 - \kappa_2)/2 -i\Delta \\
                s + (\kappa_2 - \kappa_1)/2 -i\Delta & -\sqrt{\kappa_1\kappa_2} \\
            \end{array} \right].  \nonumber \\ 
& & \hspace*{1.2em}
      \mbox{}
\end{eqnarray}
Again in the special case $\kappa_1=\kappa_2=\kappa$ and $\Delta=0$, it is 
\begin{equation}
\label{HPF}
       K(s)       
        =  \frac{1}{s+\kappa}
            \left[ \begin{array}{cc}
                -\kappa & s\\
                s & -\kappa  \\
            \end{array} \right]. 
\end{equation}
Hence the relation between $b_3$ and $\tilde{b}_4$ is given by 
\begin{equation}
	\tilde{b}_4^\dag(s^\ast) 
         = \frac{s}{s+\kappa} b_3^\dag(s^\ast)
         + \frac{-\kappa}{s+\kappa} b_4^\dag(s^\ast).\notag
\end{equation}
That is, at around $s = 0$, the cavity works as a differentiator for the 
reflected field from $b_3$ to $\tilde{b}_4$. 
Also it works as a high-pass filter with bandwidth $\kappa$; that is, the optical 
components of $b_3$ in the domain $|s|=|i\omega|\gg \kappa$ can only pass through 
the cavity. 
We also call this cavity as a MCC.


\section{Quantum feedback amplification}
\label{sec:Feedback control}

\begin{figure}[t]
\centering 
\includegraphics[width=4cm]{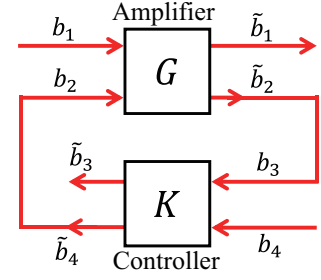}
\caption{
Feedback structure of the quantum amplifier. 
}
\label{quantum fb}
\end{figure}

In this paper we consider the general feedback-connected system shown in 
Fig.~\ref{quantum fb}, composed of the high-gain quantum phase-preserving 
amplifier $G$ and a passive system $K$. 
The feedback structure is made by 
\[
      \tilde{b}_2=b_3,~~~
      b_2=\tilde{b}_4,
\]
which are of course the same as $\tilde{b}_2^\dagger=b_3^\dagger$ and 
$b_2^\dagger=\tilde{b}_4^\dagger$. 
The entire system has the inputs $(b_1, b_4^\dagger)$ and the outputs 
$(\tilde{b}_1, \tilde{b}_3^\dagger)$. 
From Eqs.~\eqref{input output relation of NDPA} and \eqref{input output relation of MCC}, 
the input-output relation of this system is given by
\begin{eqnarray}
& & \hspace*{-1em}
      \left[ \begin{array}{c}
                \tilde{b}_1(s) \\ 
                \tilde{b}_3^\dagger(s^\ast) \\
            \end{array} \right]
          = G^{({\rm fb})}(s)
            \left[ \begin{array}{c}
                b_1(s) \\ 
                b_4^\dagger(s^\ast) \\
            \end{array} \right],
\nonumber \\ & & \hspace*{0em}
         G^{({\rm fb})}(s) =
         \left[ \begin{array}{cc}
               G_{11}^{({\rm fb})}(s) & G_{12}^{({\rm fb})}(s) \\
               G_{21}^{({\rm fb})}(s) & G_{22}^{({\rm fb})}(s) \\
             \end{array} \right],
             \label{closed system}
\end{eqnarray}
where 
\begin{align}
G_{11}^{({\rm fb})}
    =&\frac{G_{11} - K_{21}\det{[G]}}{1-K_{21}G_{22}},\label{G11fb}\\
G_{12}^{({\rm fb})}
    =&\frac{G_{12} K_{22}}{1-K_{21}G_{22}},\label{G12fb}\\
G_{21}^{({\rm fb})}
    =&\frac{G_{21} K_{11}}{1-K_{21}G_{22}},\label{G21fb}\\
G_{22}^{({\rm fb})}
    =&\frac{K_{12} + G_{22}\det{[K]}}{1-K_{21}G_{22}}\label{G22fb},
\end{align}
with $\det{[G]}= G_{11}G_{22}-G_{12}G_{21}$ and 
$\det{[K]}= K_{11}K_{22}-K_{12}K_{21}$. 
These matrix entries satisfy 
$|G_{11}^{({\rm fb})}(i\omega)|^2 - |G_{12}^{({\rm fb})}(i\omega)|^2 =1$, $\forall \omega$, etc, 
meaning that it also functions as a phase-preserving amplifier.

It was shown in \cite{Yamamoto 2016} that 
$|G_{11}^{({\rm fb})}(i\omega)| \approx 1/|K_{21}(i\omega)|$ holds in the high-gain 
amplification limit $|G_{11}(i\omega)| \rightarrow \infty$; 
because the characteristic change in the passive transfer function $K(s)$ is usually 
very small, this realizes the robust quantum amplification, which is the quantum 
analogue to the classical feedback amplification technique mentioned in the first 
paragraph in Sec.~\ref{sec:Introduction}. 
We now extend this idea to the Laplace domain. 
The point to derive the result is that, from Eq.~\eqref{G condition}, we have 
\begin{align}
& & \hspace*{-1em}
    \det{[G(i\omega)]}
        =G_{11}(i\omega)G_{22}(i\omega)
           -G_{12}(i\omega)
              \frac{G_{22}(i\omega)G_{12}^\ast(i\omega)}{G_{11}^\ast(i\omega)}
\nonumber \\ & & \hspace*{0em}
    =\frac{(|G_{11}(i\omega)|^2
        -|G_{12}(i\omega)|^2)G_{22}(i\omega)}{G_{11}^\ast(i\omega)}
    =\frac{G_{22}(i\omega)}{G_{11}^\ast(i\omega)},
\nonumber
\end{align}
and thus 
\[
      \frac{\det{[G(i\omega)]}}{G_{22}(i\omega)}
         =\frac{1}{G_{11}^\ast(i\omega)} \to 0, 
\]
in the high-gain limit $|G_{11}(i\omega)| \to \infty$. 
Also again from Eq.~\eqref{G condition}, $|G_{11}(i\omega)|=|G_{22}(i\omega)|$ and 
$|G_{12}(i\omega)|=|G_{21}(i\omega)|$ hold. Then in the same limit, 
Eq.~\eqref{G condition} leads to 
\begin{align}
    &1-\left|\frac{G_{12}(i\omega)}{G_{11}(i\omega)}\right|^2
        =\frac{1}{|G_{11}(i\omega)|^2}\to 0~~\Longrightarrow~~
    \left|\frac{G_{12}(i\omega)}{G_{22}(i\omega)}\right|\to 1,
\notag\\
    &1-\left|\frac{G_{21}(i\omega)}{G_{22}(i\omega)}\right|^2
        =\frac{1}{|G_{22}(i\omega)|^2}\to 0~~\Longrightarrow~~
    \left|\frac{G_{21}(i\omega)}{G_{22}(i\omega)}\right|\to 1\notag.
\end{align}
These are equivalent to 
\[
      \frac{G_{12}(i\omega)}{G_{22}(i\omega)} \to e^{i\theta(\omega)}, ~~
      \frac{G_{21}(i\omega)}{G_{22}(i\omega)} \to e^{i\varphi(\omega)}, 
\]
where $\theta(\omega)$ and $\varphi(\omega)$ are certain real functions of $\omega$.

We now extend the above result and assume that 
\begin{equation}
\label{G condition in s}
     \frac{\det{[G(s)]}}{G_{22}(s)} \to 0, ~~
     \frac{G_{12}(s)}{G_{22}(s)} \to 1, ~~
     \frac{G_{21}(s)}{G_{22}(s)} \to 1
\end{equation}
hold in the domain $\mathcal{D}=\{  s\in \mathbb{C}\; ;\; |G_{11}(i\omega))|\to \infty \}$. 
Moreover, we assume $G_{11}(s) = G_{22}(s)$ for all $s\in{\mathbb C}$. 
These conditions are indeed satisfied in the case of NDPA shown in 
Sec.~\ref{sec:Phase preserving linear amplifier}, for $s$ satisfying $|s|\ll \gamma$, 
where the high-gain limit is realized by taking $\gamma \to 2\lambda+0$. 
Under the above assumptions, the transfer function matrix of the entire 
closed-loop system can be approximated by 
\begin{equation}
\label{main result}
    G^{({\rm fb})}(s)= \frac{-1}{K_{21}(s)}
            \left[ \begin{array}{cc}
               1 & K_{22}(s) \\
               K_{11}(s) & \det{[K(s)]} \\
            \end{array} \right],
\end{equation}
in the domain $\mathcal{D}$. 
Hence, we now have a quantum system that, as will be proven later, generates 
several interesting and robust functionalities available in the feedback 
amplification setting.

The proof of Eq.~\eqref{main result} is as follows:
\begin{align}
    G_{11}^\textrm{(fb)}
      =&\frac{G_{11}-K_{21}\det{[G]}}{1-K_{21}G_{22}}
      =\frac{1 - K_{21}( \det{[G]}/G_{22})}{(1/G_{22})-K_{21}}\notag\\ 
      \to& -\frac{1}{K_{21}},\notag\\
      G_{12}^\textrm{(fb)}
          =&\frac{G_{12}K_{22}}{1-K_{21}G_{22}}
          =\frac{(G_{12}/G_{22})K_{22}}{(1/G_{22})-K_{21}} \to  -\frac{K_{22}}{K_{21}},\notag\\
     G_{21}^\textrm{(fb)}
     =&\frac{G_{21}K_{11}}{1-K_{21}G_{22}}
     =\frac{(G_{21}/G_{22})K_{11}}{(1/G_{22})-K_{21}} \to 
         -\frac{K_{11}}{K_{21}}.\notag\\
      G_{22}^\textrm{(fb)}
        =&\frac{K_{12}+G_{22}\det{[K]}}{1-K_{21}G_{22}}
        =\frac{(K_{12}/G_{22}) + \det{[K]}}{(1/G_{22})-K_{21}}
\notag\\
        \to & - \frac{\det{[K]}}{K_{21}}.
\notag
\end{align}
We again emphasize that Eq.~\eqref{main result} is the system only depending on 
the passive component $K$, meaning that $G^\textrm{(fb)}$ is robust against the 
characteristic change in $G$. 
Note that, for a general phase preserving amplifier which does not necessarily 
satisfy Eq.~\eqref{G condition in s} and 
$G_{11}(s) = G_{22}(s)~\forall s\in{\mathbb C}$, the resulting closed-loop system 
in the high-gain limit may still contain some components of $G$. 
Hence, following the convention of the classical feedback amplification theory, 
we call $G$ satisfying these assumptions the {\it ideal quantum op-amp}.


\section{Stability analysis method}
\label{sec:Stability analysis method}

\begin{figure}[t]
\centering 
\includegraphics[width=5.5cm]{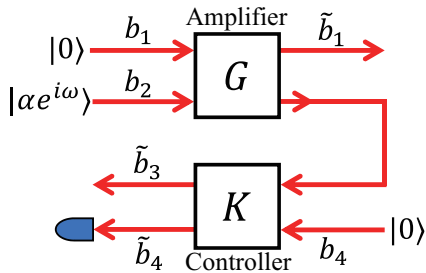}
\caption{
System structure for the stability test of the closed-loop system
}
\label{stability analysis}
\end{figure}

From an engineering viewpoint, it is important to guarantee the stability of 
the entire controlled system before activating it (more precisely, before closing the loop for control). 
In the classical case the seminal Nyquist method \cite{feedback systems}
is often used for this purpose. Here we show the quantum version of this method, 
particularly for the quantum 
feedback-controlled system with transfer function matrix \eqref{closed system}; 
note that, hence, the stability must be guaranteed for the system with finite 
amplification gain.

Let us represent the matrix entries of $G(s)$ and $K(s)$ as 
$G_{ij}(s)=g_{ij}(s)/g(s)$ and $K_{ij}(s)=k_{ij}(s)/k(s)$, respectively, where 
$g(s), g_{ij}(s), k(s)$, and $k_{ij}(s)$ are the polynomial functions. 
Then, it is easy to see that $G^{({\rm fb})}(s)$ has the following form: 
\begin{equation}
     G^{({\rm fb})}(s)
         = \frac{1}{g^2(s) k^2(s) \{1-K_{21}(s)G_{22}(s)\}}
            \left[ \begin{array}{cc}
             \star & \star \\
             \star & \star \\
            \end{array} \right], \notag
\end{equation}
where for simplicity the matrix entries, the polynomial functions denoted by 
$\star$, are not shown. 
Here we assume that the original systems $G(s)$ and $K(s)$ are stable; 
then because $k(s)$ and $g(s)$ are stable polynomial functions (meaning that 
the zeros of $k(s)$ and $g(s)$ lie in the left side plane in ${\mathbb C}$), 
the stability of the closed-loop system is completely characterized by the 
zeros of $1-K_{21}(s)G_{22}(s)$.

We can now apply the classical Nyquist method to test the stability of this 
closed-loop system. 
As in the classical case let us define the open-loop transfer function:
\[
       L(s) = -K_{21}(s)G_{22}(s).
\]
Then, from the Nyquist theorem, the simplest stability criterion is as follows. 
If the point $-1$ lies outside the Nyquist plot (i.e., the trajectory of 
$L(i\omega)$ for $\omega \in (-\infty, +\infty)$ in the complex plane), then 
the closed-loop system is stable; 
otherwise, it is unstable. 
The point is that this stability test can be carried out for an open-loop system 
illustrated in Fig.~\ref{stability analysis}, which is constructed via simply 
cascading the amplifier and the controller. 
In fact the input-output relation of this open-loop system is given by 
\begin{align}
      \left[ \begin{array}{c}
                \tilde{b}_1(s) \\ 
                \tilde{b}_3^\dagger(s^\ast) \\
                \tilde{b}_4^\dagger(s^\ast) \\
            \end{array} \right]
        =&G^\textrm{(open)}(s)
            \left[ \begin{array}{c}
                b_1(s) \\ 
                b_2^\dagger(s^\ast) \\
                b_4^\dagger(s^\ast) \\
            \end{array} \right],\notag\\
       G^\textrm{(open)}(s)=& \left[ \begin{array}{ccc}
             G_{11}(s) & G_{12}(s) & 0 \\
             K_{11}(s)G_{21}(s) & K_{11}(s)G_{22}(s) & K_{12}(s) \\
             K_{21}(s)G_{21}(s) & K_{21}(s)G_{22}(s) & K_{22}(s) \\
            \end{array} \right].\notag
\end{align}
Therefore, the Nyquist plot can be obtained by setting $b_1$ and $b_4$ to 
the vacuum fields and injecting the coherent field $\ket{\alpha e^{i\omega}}$ 
in the $b_2$ port with several frequencies $\omega$. 
In fact, measuring the amplitude of the output field $\tilde{b}_4$ gives us 
the Nyquist plot in the form 
$L(i\omega)=-\mean{\tilde{b}_4^\dagger(-i\omega)}/\alpha^*$. 
Note that the measurement result of $\tilde{b}_4$ must be probabilistic, and 
hence the Nyquist plot constitutes a ``band" with variance 
$\mean{|\Delta\tilde{b}_4(i\omega)|^2}$, meaning that the stability margin 
should be taken into account.


\section{Functionalities 1: Quantum PID}
\label{sec:Functionalities}

We now start describing several functionalities realized by the developed 
quantum feedback amplification method. 
The first functionality is the quantum PID \cite{Ang 2005}. 
That is, we show that, via the proper choice of the system $K$, the ideal closed-loop system 
\eqref{main result} functions as a differentiator (D) or integrator (I) on the input itinerant field $b_1$; 
hence together with the proportional component (P), which simply attenuates or amplifies the 
amplitude of the input, now P, I, and D components are provided to us. 
These three are clearly the most basic components involved in almost all electrical circuits and 
used for constructing several useful systems such as a PID feedback controller and an analogue 
computer. 
In fact, there have been proposed a classical PID control for quantum systems 
\cite{Gough kalman pid, q pi control}; the fully-quantum PID controller constructed using 
feedback amplification may have some advantages over the classical one, but we leave 
this analysis for future work. 
Instead, we will show a simple application of the quantum integrator at the end of this section.


\subsection{Differentiator}

\begin{figure}[t]
\begin{minipage}{\hsize}
\centering
\centerline{\hspace{2em}\normalsize(a)}
\hspace{2em}\includegraphics[width=0.8\columnwidth]{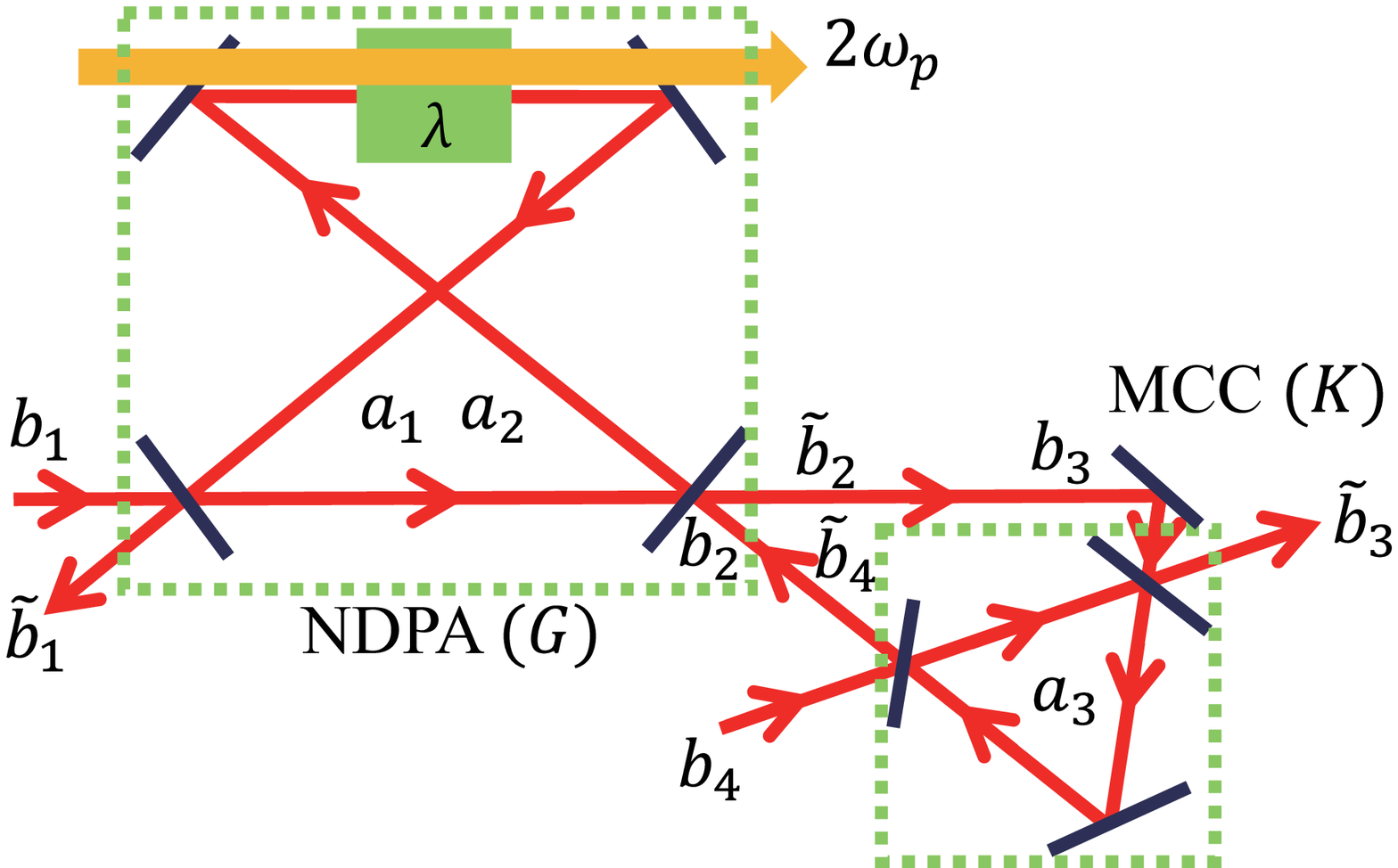}
\vspace{1em}
\label{differentiator}
\end{minipage}\\
\begin{minipage}{0.65\hsize}
\centering
\centerline{\hspace{2em}\normalsize(b)}
\includegraphics[width=\columnwidth]{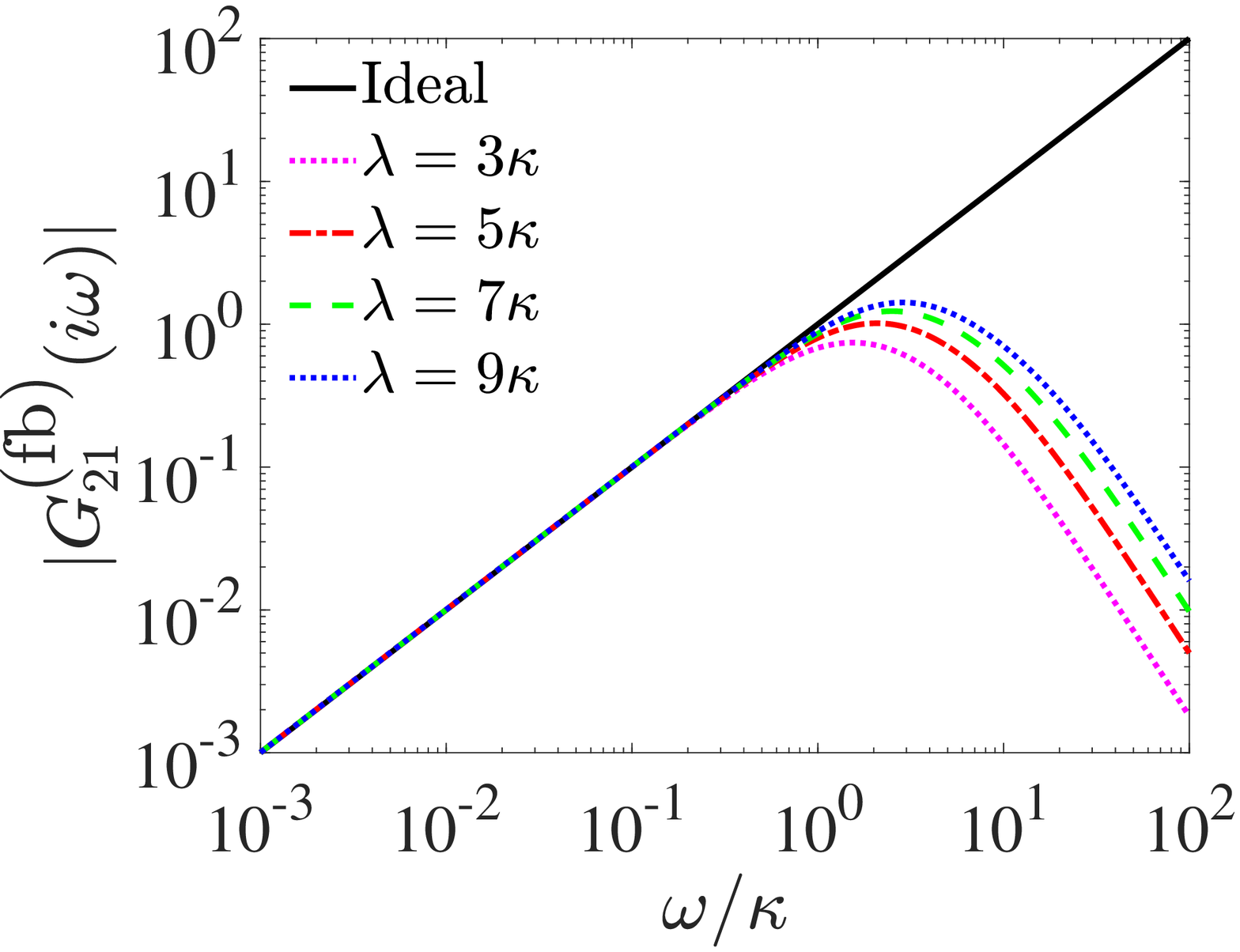}
\label{gain_of_differentiator}
\end{minipage}
\caption{
(a) Optical realization of the quantum differentiator. 
(b) Gain plot of the actual transfer function \eqref{G21fb} and its ideal 
limit $|i\omega/\kappa|$ for the quantum differentiator.}
\label{quantum differentiator (a)(b)}
\end{figure}

Let us take the symmetric cavity \eqref{LPF} as the controller $K(s)$. 
In this case, the transfer function of the ideal closed-loop system 
\eqref{main result} is given by %
\[
       G^{({\rm fb})}(s)       
        =  \frac{1}{\kappa}
            \left[ \begin{array}{cc}
                s + \kappa & s\\
                s & s -\kappa  \\
            \end{array} \right]. 
\]
Hence from Eq.~\eqref{closed system}, the output $\tilde{b}_3^\dagger(s)$ is given by 
\[
     \tilde{b}_3^\dagger(s^\ast) 
         = \frac{s}{\kappa} b_1(s)
         + \frac{s-\kappa}{\kappa} b_4^\dagger(s^\ast), 
\]
or in the time-domain it is 
\begin{equation}
            \tilde{b}_3^\dagger(t) 
         = \frac{1}{\kappa}\frac{d}{dt} b_1(t)
         + \frac{1}{\kappa} \frac{d}{dt} b_4^\dagger(t) - b_4^\dagger(t),\notag
\end{equation}
meaning that the closed-loop system works as a differentiator for the 
itinerant field $b_1(t)$.

As discussed in Sec.~III, the approximation is valid only in a specific 
$s$-region such that the high-gain limit is effective. 
To show a concrete example of this region, 
we study a feedback controlled system composed 
of the optical NDPA and the control cavity, depicted in 
Fig.~\ref{quantum differentiator (a)(b)}(a). 
Recall that, in the case of NDPA, the high-gain limit is achieved in the 
regime $|s|\ll \gamma \approx 2\lambda$, which becomes wider as $\lambda$ 
increases. 
Actually this can be seen in Fig.~\ref{quantum differentiator (a)(b)}(b), 
showing the gain plot of the transfer function \eqref{G21fb} of this optical 
system with parameters $\gamma=2.01\lambda$ and its high-gain limit 
$|i\omega/\kappa|$; 
that is, the frequency range such that this controlled system effectively 
approximates the ideal differentiator becomes wider as $\lambda$ gets bigger.

\begin{figure}[h]
\centering
\leftline{\hspace{2.5cm}\normalsize(a)\hspace{3.75cm}\normalsize(b)}
\includegraphics[width=\columnwidth]{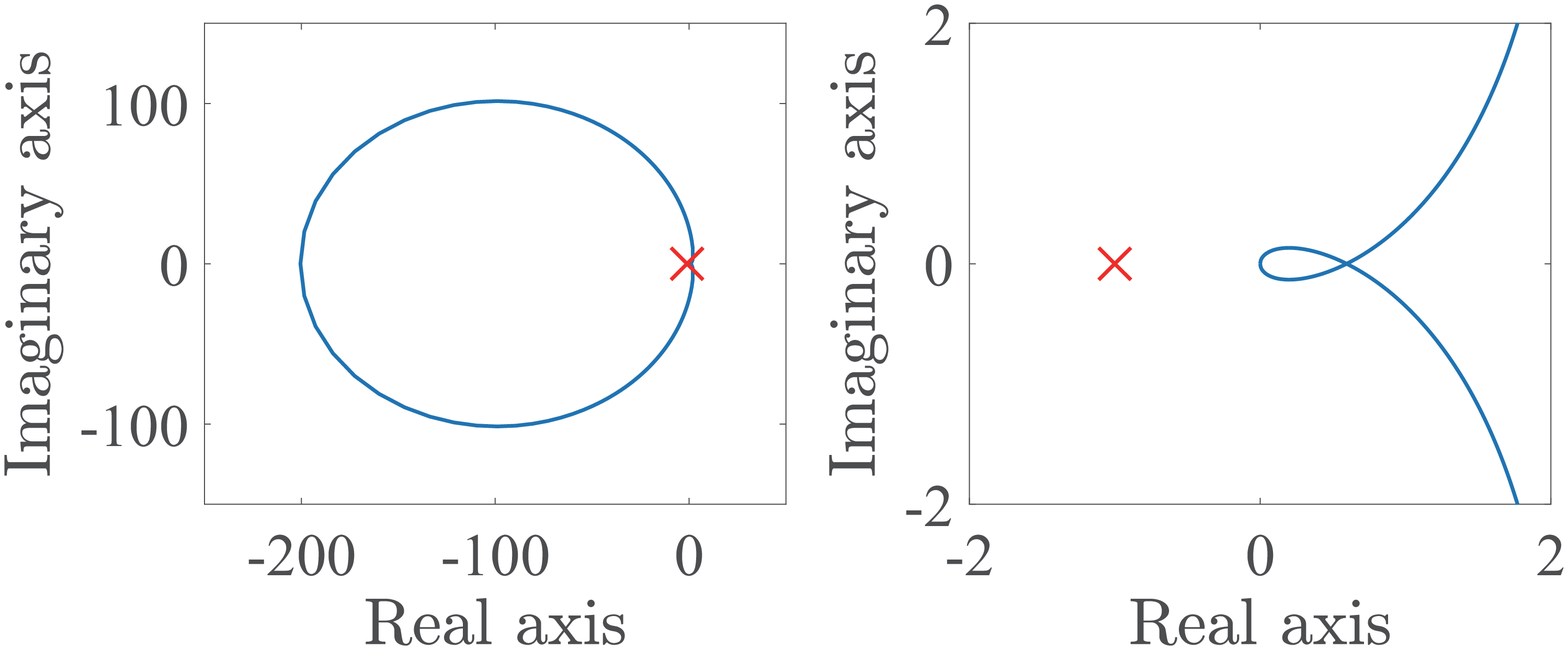}
\caption{
(a) Nyquist plot of the quantum differentiator, where the parameters are set as 
$\kappa=1$, $\lambda=2\kappa$, and $\gamma=2.01\lambda$. 
Figure (b) is a zoom-up of (a) at around $s=0$.}
\label{Nyquist differentiator}
\end{figure}

Note that, as in the classical case, the differentiator itself is an unstable 
system, and thus this system should be used together with other components 
such that the entire system is stable. 
This instability can be readily seen using the method addressed in 
Sec.~\ref{sec:Stability analysis method}; 
the open-loop transfer function in this case is 
\[
       L(s)=\frac{\kappa(s^2-\gamma^2/4-\lambda^2)}
                      {(s+\kappa)(s^2+\gamma s+\gamma^2/4-\lambda^2)}, 
\]
and the Nyquist's plot is given by Figs.~\ref{Nyquist differentiator}(a) and 
(b), showing that the point $-1$ lies inside the trajectory of $L(i\omega)$ 
and thus the system is unstable. 
Note that the actual Nyquist's plot fluctuates along the curve shown in the 
figure, with variance $\mean{|\Delta\tilde{b}_4(i\omega)|^2}$.


\subsection{Integrator}
\label{sec:Integrator}

\begin{figure*}[t]
\begin{minipage}{\hsize}
\centering
\leftline{\hspace{4.5cm}\normalsize(a)\hspace{8cm}\normalsize(b)}
\includegraphics[width=0.4\columnwidth]{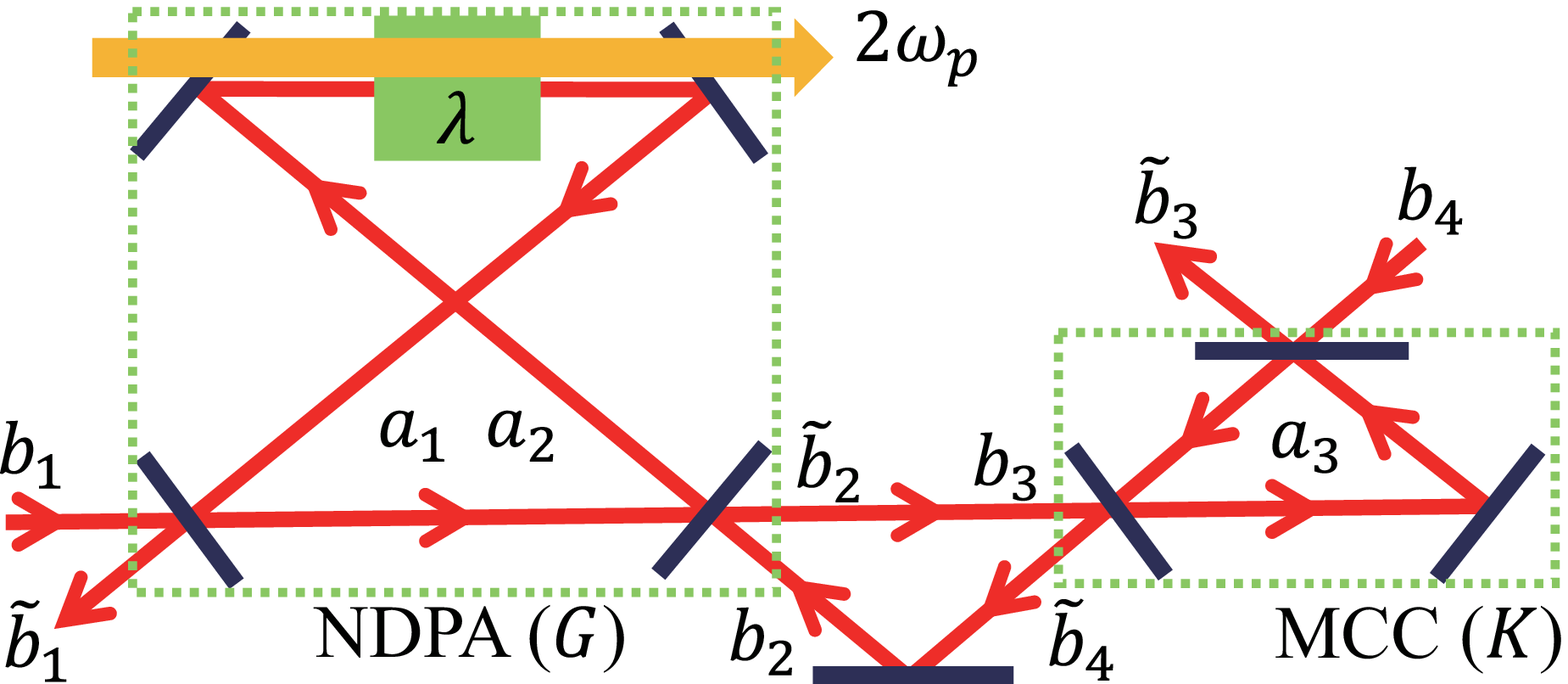}
\hspace{3em}
\includegraphics[width=0.4\columnwidth]{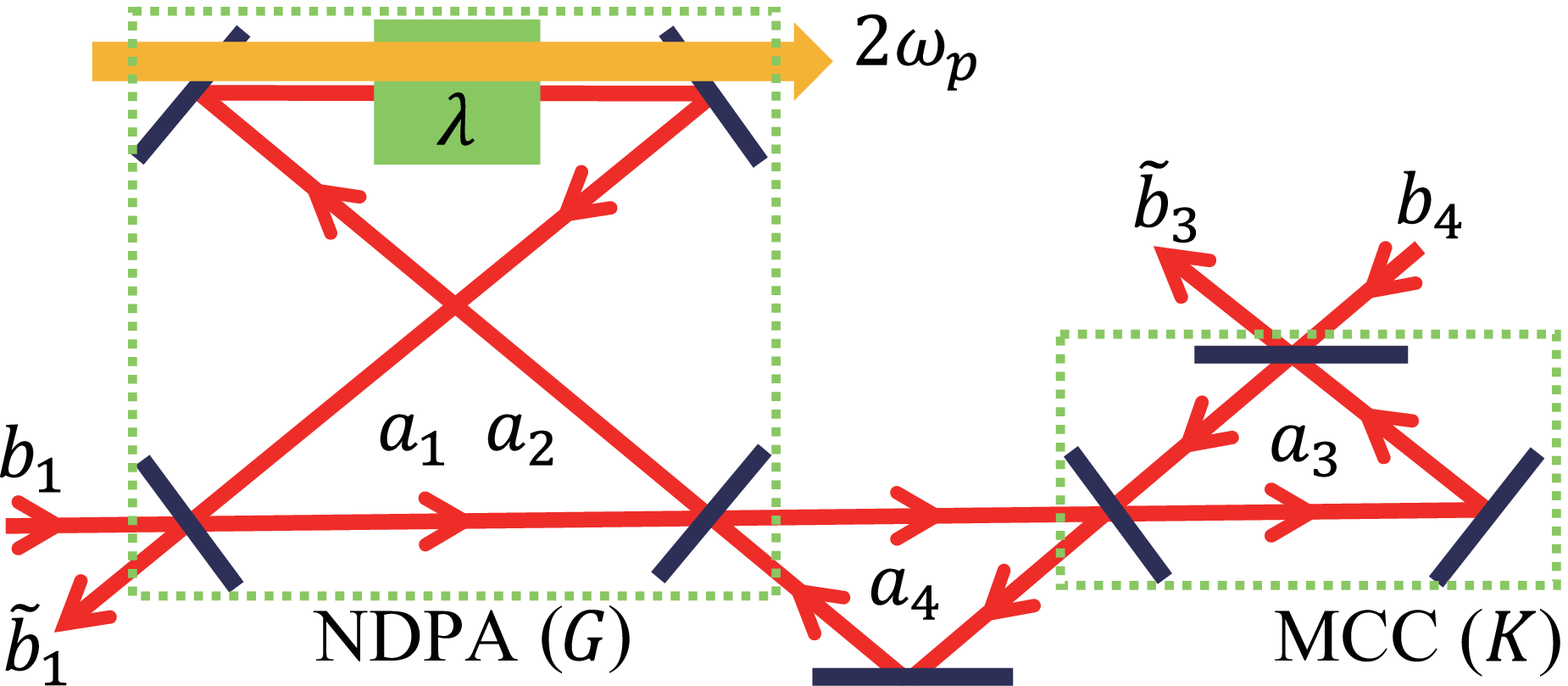}
\end{minipage}\\
\begin{minipage}{\hsize}
\vspace{2em}
\end{minipage}\\
\begin{minipage}{0.5\hsize}
\centering
\vspace{-2em}\
\centerline{\hspace{-2cm}\normalsize(c)}
\includegraphics[width=\columnwidth]{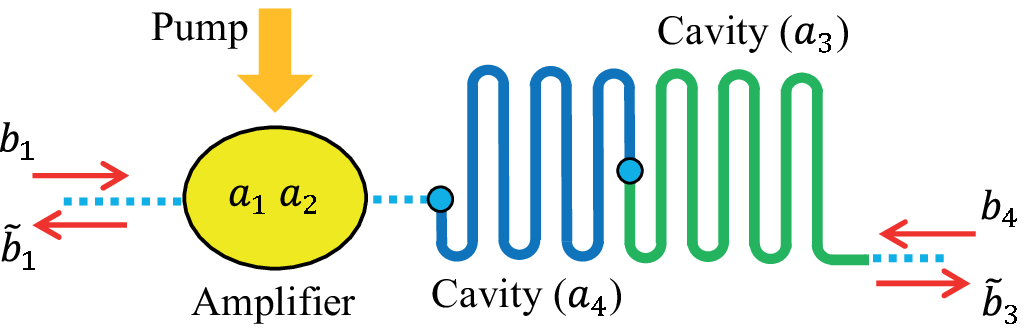}
\end{minipage}
\hspace{-2em}
\begin{minipage}{0.4\hsize}
\centerline{\hspace{1cm}\normalsize(d)}
\includegraphics[width=0.7\columnwidth]{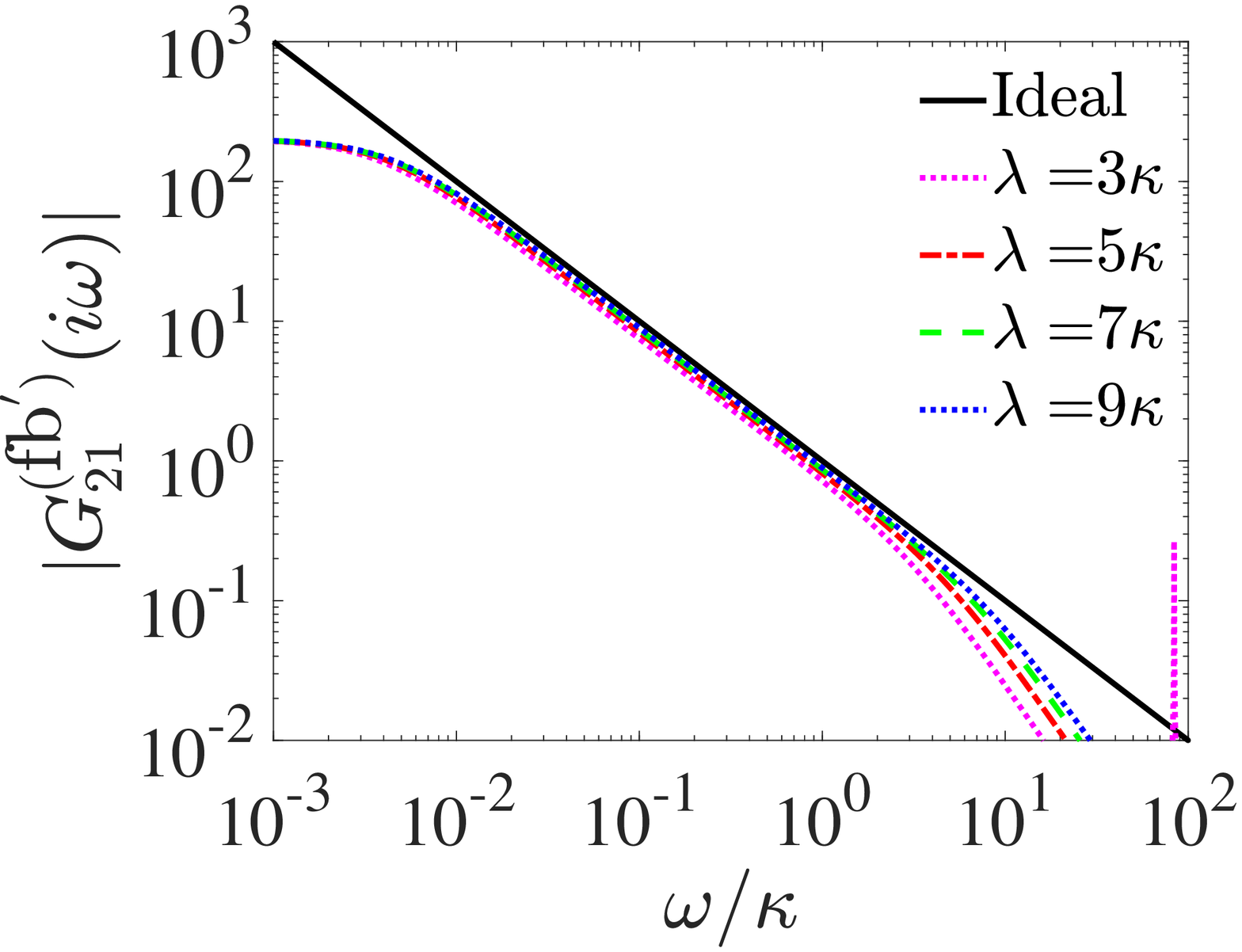}
\end{minipage}
\caption{
(a) Optical realization of the quantum integrator with input $b_1$ and 
output $\tilde{b}_3$. 
(b) Modified model for the quantum integrator where the loop between the 
NDPA and MCC is regarded as a cavity with the mode $a_4$. 
(c) Microwave realization of the quantum integrator.
(d) Gain plot of the actual transfer function $|G_{21}^{\rm (fb')}(i\omega)|$ 
in Eq.~\eqref{eq:integrator G21 cavity} and its ideal limit 
$|\kappa/i \omega|$. 
The spike around $\omega/\kappa\approx10^2$ is due to invalid approximation 
in the high frequency region.}
\label{integrator (a)(b)(c)(d)}
\end{figure*}

Next we take the high-pass filtering cavity \eqref{HPF} as the controller, 
where in this case the reflected field is fed back to the amplifier, as 
shown in Fig.~\ref{integrator (a)(b)(c)(d)}(a) in optics case. 
Then the transfer function matrix \eqref{main result} of the ideal closed-loop 
system is given by 
\[
       G^{({\rm fb})}(s)       
        =  \frac{1}{s}
            \left[ \begin{array}{cc}
                - s - \kappa & \kappa\\
                \kappa & s -\kappa  \\
            \end{array} \right]. 
\]
Hence from Eq.~\eqref{closed system} the output $\tilde{b}_3^\dagger(s^\ast)$ 
is connected to the input $b_1(s)$ as
\[
     \tilde{b}_3^\dagger(s^\ast) 
         = \frac{\kappa}{s} b_1(s)
         + \frac{s-\kappa}{s} b_4^\dagger(s^\ast), 
\]
or in the time-domain it is 
\begin{equation}
         \tilde{b}_3^\dagger(t) 
         = \kappa \int_0^t b_1(\tau)d\tau 
         + b_4^\dag(t)
         - \kappa \int_0^t b_4^\dagger(\tau)d\tau .\label{eq: integrator time domain}
\end{equation}
This means that in a specific $s$-region where the high-gain limit of the 
amplifier is effective ($|s|\ll \gamma$ in the NDPA case, as discussed in 
Sec.~\ref{sec:Feedback control}), the closed-loop system works as an 
integrator for the itinerant field $b_1(t)$. 
Note again that the feedback controlled system can approximate the 
integrator in the low-frequency regime, while the 
standard non-controlled cavity \eqref{passive cavity input output relation 1} 
can do the same task in a high-frequency regime $|i\omega|\gg\kappa$; 
in this sense the integrator shown here is the functionality realized only via the 
feedback amplification method.

Now, unlike the differentiator, the integrator forms a circulating field 
in the feedback loop between the amplifier and the controller cavity. 
Therefore, we regard this loop as another cavity with mode $a_4$, 
as depicted in Fig.~\ref{integrator (a)(b)(c)(d)}(b) in optics case
and Fig.~\ref{integrator (a)(b)(c)(d)}(c) in a microwave system case, 
which we call the loop cavity. 
In fact, for the model depicted in Fig.~\ref{integrator (a)(b)(c)(d)}(a) where 
$b_2$, $\tilde{b}_2$, $b_3$, and $\tilde{b}_4$ are treated as the itinerant 
fields, they violate the Ito rule such as $dB_2(t) dB_2^\dag(t)=dt$ 
with $B_2(t)=\int_0^t b_2(\tau)d\tau$ \cite{Gardiner Book, Hudson 1984}. 
In what follows we show that this modified model still maintains the 
functionality of integration. 
More precisely, we will show that the transfer function from $b_1$ to $\tilde{b}_3^\dag$ 
in Fig.~\ref{integrator (a)(b)(c)(d)}(b) or Fig.~\ref{integrator (a)(b)(c)(d)}(c) 
approaches to $\kappa/s$ in the same high-gain limit.

First, the Hamiltonian of the system is given by 
\begin{align}
      H_\textrm{fbamp}
       = &\sum_{k=1}^{4}\hbar\omega_ka^\dag_ka_k
       +i\hbar\lambda(a_1^\dagger a_2^\dagger e^{-2i\omega_pt}-a_1a_2e^{2i\omega_pt})
\notag\\
    & \hspace{-1em} +\hbar g_{24}(a_2^\dagger a_4+a_2a_4^\dagger)
                  +\hbar g_{34}(a_3^\dagger a_4+a_3a_4^\dagger),
\label{three cascaded cavities}
\end{align}
where $\omega_3$ and $\omega_4$ are the resonant frequencies of $a_3$ and $a_4$, 
respectively. 
$g_{24}$ ($g_{34}$) describes the coupling strength 
between $a_2$ and $a_4$ ($a_3$ and $a_4$), which are given by 
\begin{equation}
         g_{24}=\sqrt{c\gamma/L_4}, \quad 
         g_{34}=\sqrt{c\kappa/L_4}, 
\label{eq:g24g34definition integrator}
\end{equation}
with $L_4$ the round trip length of the loop cavity and $c=3\times10^8$ m/s 
the speed of light. 
Here we assume $\omega_k=\omega_p$ ($k=$1, \ldots, 4). 
Together with the coupling to the external fields, we find that, in the rotating frame at 
frequency $\omega_p$, the dynamical equations are given by
\begin{align}
    & \dot{a}_1= -\frac{\gamma}{2}a_1+\lambda a_2^\dag-\sqrt{\gamma}b_1, ~~
        \dot{a}_2^\dag = \lambda a_1+ig_{24}a_4^\dag,
\notag \\
    & \dot{a}_3^\dag = -\frac{\kappa}{2}a_3^\dag+ig_{34}a_4^\dag-\sqrt{\kappa}b_4^\dag, ~~
        \dot{a}_4^\dag = ig_{24}a_2^\dag+ig_{34}a_3^\dag,
\notag \\
     & \tilde{b}_1 = \sqrt{\gamma}a_1 + b_1, ~~
     \tilde{b}_3^\dag = \sqrt{\kappa}a_3^\dag + b_4^\dag.
\label{three cavities dynamics}
\end{align}
The input-output equation of this system in the Laplace domain is of the form 
\[
      \left[ \begin{array}{c}
                \tilde{b}_1(s) \\ 
                \tilde{b}_3^\dagger(s^\ast) \\
            \end{array} \right]
         = \left[ \begin{array}{cc}
               G_{11}^{({\rm fb'})}(s) & G_{12}^{({\rm fb'})}(s) \\
               G_{21}^{({\rm fb'})}(s) & G_{22}^{({\rm fb'})}(s) \\
             \end{array} \right]
           \left[ \begin{array}{c}
                 b_1(s) \\ 
                 b_4^\dagger(s^\ast) \\
             \end{array} \right], 
\]
where particularly $G_{21}^{({\rm fb'})}(s)$ is given by
\begin{equation}
      G_{21}^{({\rm fb'})}(s)
        =\frac{\alpha_0}{s^4+\beta_3 s^3+ \beta_2 s^2+\beta_1 s +\beta_0},
\label{eq:integrator G21 cavity}
\end{equation}
with 
\begin{align}
     \alpha_0=&\sqrt{\gamma\kappa}\lambda g_{24}g_{34}, ~~
     \beta_0=\gamma\kappa g_{24}^2/4-\lambda^2g_{34}^2, 
\notag\\
      \beta_1=&(\gamma g_{24}^2+\kappa g_{24}^2+\gamma g_{34}^2-\kappa\lambda^2)/2,
\notag\\
       \beta_2=&\gamma\kappa/4-\lambda^2+g_{24}^2+g_{34}^2, ~~
       \beta_3=(\gamma+\kappa)/2. 
\notag
\end{align}
In the limit $\gamma\to2\lambda+0$, together with 
Eq.~\eqref{eq:g24g34definition integrator}, the above coefficients are approximated as 
\begin{align}
    \alpha_0\approx & 2c\kappa\lambda^2/L_4, ~~
    \beta_0\approx 0, 
\notag \\
    \beta_1\approx & 2c\lambda(\kappa+\lambda)/L_4-\kappa\lambda^2/2, 
\notag \\
     \beta_2\approx & \kappa\lambda/2-\lambda^2+c(\kappa+2\lambda)/L_4, ~~
      \beta_3\approx \kappa/2+\lambda. 
\notag
\end{align}
Furthermore, we assume $|s|\ll\gamma$, so that the higher-order term of $s$ can 
be neglected. 
Then the transfer function \eqref{eq:integrator G21 cavity} can be approximated by
\begin{align}
      G_{21}^{({\rm fb'})}(s)
       \approx & \frac{\alpha_0}{\beta_1s}
       \approx
         \frac{2c\kappa\lambda^2/L_4}
                { \left\{2c\lambda(\kappa+\lambda)/L_4-\kappa\lambda^2/2 \right\} s}
\notag\\
        =&\frac{\kappa}
              {\dis \left(1+\frac{\kappa}{\lambda}- \frac{L_4\kappa}{4c}\right)s}.\notag
\end{align}
Thus, if $\kappa\ll\lambda$ and $L_4\kappa/(4c)\ll1$, 
this system becomes the integrator which we wish to obtain:
\begin{equation}
    G_{21}^{({\rm fb'})}(s)\approx\frac{\kappa}{s}.
\notag
\end{equation}

In Fig.~\ref{integrator (a)(b)(c)(d)}(d), the solid black line shows  
the ideal gain plot of the integrator, (i. e., $|\kappa/i\omega|$), 
while the dotted lines represent 
$|G_{21}^{({\rm fb'})}(i\omega)|$ in Eq.~\eqref{eq:integrator G21 cavity} 
with parameters $\gamma=2.01\lambda$, $c/L_4=10^3\kappa$, and 
$\lambda=n\kappa$ ($n=3, 5, 7, 9$). 
Clearly, $|G_{21}^{({\rm fb'})}(i\omega)|$ well approximates $|\kappa/i\omega|$ 
in a specific $s$ region where the high-gain limit of the NDPA is effective, 
which is now given by $|s|\ll \gamma \approx 2\lambda$. 
Hence, $\lambda$ should be relatively large to guarantee that the integrator 
works in a wider region in $s$; this can be actually seen in the figure, 
although making $\lambda$ large does not make a big difference in the 
parameter regime considered here. 
However, the approximation is not valid in the frequency domain where 
$\omega / \kappa < 10^{-2}$. 
This is because, if $s=0$, the transfer function in Eq.~\eqref{eq:integrator G21 cavity} is 
$G_{21}^\textrm{(fb')}(0)=\alpha_0/\beta_0$, whereas the transfer function of the integrator 
goes towards infinity if $s\to0$. 
Therefore, the approximation is valid in the relatively low frequency domain where $s^4, s^3, s^2$ 
can be ignored and where $|\beta_1 s| \gg |\beta_0|$ ($\Leftrightarrow |s| \gg |\beta_0/\beta_1|$).

The stability of the modified model depicted in 
Figs.~\ref{integrator (a)(b)(c)(d)}(b) and \ref{integrator (a)(b)(c)(d)}(c) 
cannot be investigated via the 
stability test discussed in Sec.~\ref{sec:Stability analysis method}, 
which can only be applied to the case where the feedback loop does not 
form a cavity. 
Therefore, instead of the Nyquist's theorem, we use the 
{\it Routh-Hurwitz method} \cite{routh-hurwitz}. 
In our case, the system is stable if and only if every root of the 
characteristic polynomial function in the denominator in 
Eq.~\eqref{eq:integrator G21 cavity} has a negative real part; 
the Routh-Hurwirz method systematically leads to the stability condition 
as follows: 
\begin{equation}
    \beta_3>0,~~\frac{\beta_2\beta_3-\beta_1}{\beta_3}>0,~~
    \frac{\beta_3^2\beta_0}{\beta_1-\beta_2\beta_3}+\beta_1>0.
\notag
\end{equation}
Note that $\beta_3>0$ is already satisfied.


\subsection{Application to qubit detection}

Here we give an application of the integrator, which can be used in a 
stand-alone fashion unlike the differentiator. 
The system of interest (not the feedback controlled system) is a 
qubit that is dissipatively coupled to the external itinerant field 
$b_0(t)$, such as a transmon qubit coupled to a superconducting 
resonator; 
the Langevin equation of the system variable $\sigma_x(t)$ is given by 
\[
       \frac{d}{dt}\sigma_x(t) = -\frac{\Gamma}{2}\sigma_x(t)
                 + \sqrt{\Gamma}\sigma_z(t) \{b_0(t)+b^\dagger_0(t)\}, 
\]
where $\Gamma$ is the strength of the dissipative coupling 
\cite{Gardiner Book}. 
The output field is given by $b_1(t)=\sqrt{\Gamma}\sigma_-(t) + b_0(t)$; 
the quadrature $q_1(t)=\{b_1(t)+b_1^\dagger(t)\}/\sqrt{2}$ thus follows 
\[
     q_1(t) = \sqrt{\frac{\Gamma}{2}}\sigma_x(t) 
                + \frac{b_0(t)+b^\dagger_0(t)}{\sqrt{2}}. 
\]
Now the field state is set to the vacuum $\ket{0}_F$. 
Then we find 
\[
     \mbox{}_F\bra{0}q_1(t)\ket{0}_F 
       = \sqrt{\frac{\Gamma}{2}}e^{-\Gamma t/2}\sigma_x,
\]
where for a system-field operator $X$, $\mbox{}_F\bra{0}X\ket{0}_F$ 
represents an operator living in the system Hilbert space. 
This means that, for a very short time interval $\Gamma t\ll 1$, 
the above operator is approximated as 
$\mbox{}_F\bra{0}q_1(t)\ket{0}_F \approx \sqrt{\Gamma/2}\sigma_x$, 
which thus takes $\pm \sqrt{\Gamma/2}$ when measuring it. 
In other words, to measure the qubit state we need a high-speed detector.

Using the integrator changes this condition. 
Let us place the ideal integrator having the input-output relation 
\eqref{eq: integrator time domain} along the output field of the qubit. 
That is, the output $b_1(t)$ is taken as the input to the 
integrator, and we measure its output $\tilde{b}_3(t)$. 
The quadrature 
$\tilde{q}_3(t)=\{\tilde{b}_3(t)+\tilde{b}_3^\dagger(t)\}/\sqrt{2}$ then 
satisfies 
\begin{eqnarray*}
& & \hspace*{-1em}
     \mbox{}_F\bra{0}\tilde{q}_3(t)\ket{0}_F 
       = \mbox{}_F\bra{0}
                   \Big(\kappa \int_0^t \frac{b_1(\tau)+b_1^\dagger(\tau)}{\sqrt{2}} d\tau \Big)\ket{0}_F
\nonumber \\ & & \hspace*{4.9em}
       = \kappa\sqrt{\frac{2}{\Gamma}}(1-e^{-\Gamma t/2})\sigma_x, 
\end{eqnarray*}
where $b_4$ is set to be a vacuum field. 
Therefore, in the long time limit $\Gamma t \gg 1$, this output 
itinerant field becomes 
\[
          \mbox{}_F\bra{0}\tilde{q}_3(t)\ket{0}_F
          \approx \kappa\sqrt{\frac{2}{\Gamma}}\sigma_x. 
\]
Hence the measurement result is $\pm \kappa\sqrt{2/\Gamma}$, 
which equals to $\pm \sqrt{\Gamma/2}$ by choosing the integrator 
parameter as $\kappa=\Gamma/2$. 
This means that, even if the given detector is slow, the integrator assist 
it to capture the same amount of measurement signal as that obtained by a fast detector.


\section{Functionalities 2: Self oscillation}
\label{sec.Self oscillation and frequency conversion}

Self-oscillation is also an important functionality realized with the 
feedback amplification method, which is indeed widely used in a variety 
of engineering scenes. To realize a sustained oscillation, of course, 
some nonlinearities such as a voltage saturation are necessary to be involved, 
but here we only focus on the linear part; a possible practical application of 
the proposed method which combines a nonlinear component will be presented in a 
future work.

Let us consider the cavity \eqref{asymmetric HPF} with 
$\kappa_1 = \kappa_2 = \kappa$. 
Then the transfer function of the ideal closed-loop system, 
Eq.~\eqref{main result}, is given by 
\[
       G^{({\rm fb})}(s)       
        =  \frac{1}{s-i\Delta}
            \left[ \begin{array}{cc}
                 s + \kappa-i\Delta & -\kappa  \\
                \kappa   &  s - \kappa - i\Delta \\
            \end{array} \right]. 
\]
Therefore the output $\tilde{b}_3$ is given by 
\[
    \tilde{b}_3^\dagger(s^\ast) 
         = \frac{ \kappa}{ s-i\Delta } b_1(s)
           + \frac{ s-\kappa-i\Delta}{ s-i\Delta } b^\dagger_4(s^\ast). 
\]
Because the pole is on the imaginary axis, this represents a self-oscillation 
of $\tilde{b}_3$. 
In fact, if both $b_1$ and $b_4$ are set to the vacuum and 
$\mean{\tilde{b}_3(0)}\neq 0$, then in the time domain $\tilde{b}_3$ satisfies 
\begin{equation}
\label{eq:a5 self oscillation}
       \mean{\tilde{b}_3(t)}=e^{-i\Delta t}\mean{\tilde{b}_3(0)},
\end{equation}
hence it oscillates with frequency $-\Delta$ (in the rotating-frame). 
Also, the spectral broadening of this oscillation can be seen 
from 
\begin{equation}
    \frac{1}{2\pi}\int_{-\infty}^\infty
    \mean{\tilde{b}_3^\dag(-i\omega) \tilde{b}_3(-i\omega')}d\omega' 
          = \frac{ \kappa^2}{ (\omega - \Delta)^2 }\notag.
\end{equation}
In practice, the cavity parameter $\kappa_1-\kappa_2$ is set to a small positive number, 
which makes the system oscillating almost with a fixed frequency yet with growing amplitude; 
but the amplitude is saturated by some dissipative nonlinearities, and as a result 
a sustained oscillation called the quantum limit cycle can be realized. 
As in the classical case, there may be several applications of this functionality, e.g., 
a quantum memory \cite{Mabuchi 15} and synchronization for spectroscopy \cite{Xu Holland 2015}; 
also see Ref.~\cite{Kato 2019} for the general semi-classical method for analyzing the quantum 
limit cycle and synchronization.

\begin{figure}[tbp]
\centering\includegraphics[width=0.93\columnwidth]{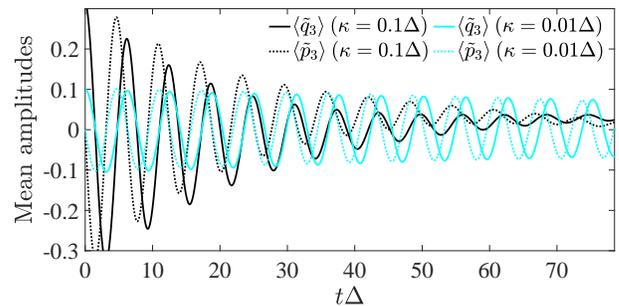}
\caption{
Time evolution of the quadratures of $\tilde{b}_3$. 
Note that the amplitudes will eventually diverge due to instability of the self-oscillator, which is not 
illustrated in the figure; in practice, such divergence is suppressed via some nonlinearity.
}
\label{fig:a5_self_oscillation}
\end{figure}

Now we show the dynamics of the entire closed loop system in nearly 
the ideal amplification limit. In the optics case, the realization of 
the self-oscillator is very similar to the integrator shown in 
Fig.~\ref{integrator (a)(b)(c)(d)}(a). 
The only difference between the self-oscillator and the integrator is 
that the detuning of the MCC is not zero for the case of self-oscillator, 
while it is zero for the integrator. 
Also as in the integrator, the self-oscillator forms a loop cavity between 
the NDPA and MCC, and thus it should be modeled as the system shown in 
Fig.~\ref{integrator (a)(b)(c)(d)}(b) or Fig.~\ref{integrator (a)(b)(c)(d)}(c) 
with non-zero detuning $\Delta$ in the mode $a_3$. 
Then the entire dynamical equation of the self-oscillator are the same as those in 
Eqs.~\eqref{three cavities dynamics} except that $i\Delta a_3^\dag$ is added 
to the right hand side of the equation that has a term $\dot{a}_3^\dag$ in the left hand side. Here, 
if $\mean{b_1(t)}=\mean{b_4^\dag(t)}=0$ $\forall t$, the mean dynamics is 
\begin{equation}
    \frac{d}{dt}\left[
\begin{array}{c}
\mean{a_1}  \\
\mean{ a_2^\dag} \\
\mean{ a_3^\dag} \\
 \mean{a_4^\dag}
\end{array}
\right]=A_\textrm{osc}
\left[
\begin{array}{c}
\mean{a_1}  \\
\mean{ a_2^\dag} \\
\mean{ a_3^\dag} \\
 \mean{a_4^\dag}
\end{array}
\right],\quad 
\mean{\tilde{b}_3^\dag}=\sqrt{\kappa}\mean{a_3^\dag},\notag
\end{equation}
where
\begin{equation}
    A_\textrm{osc}=\left[
\begin{array}{cccc}
-\gamma/2 &\lambda &0 &0  \\
\lambda &0 &0 &ig_{24}  \\
 0& 0&-\kappa/2+i\Delta &ig_{34}  \\
 0& ig_{24}&ig_{34} &0 
\end{array}
\right].\notag
\end{equation}
Figure \ref{fig:a5_self_oscillation} shows the mean time evolution of the quadratures of 
$\tilde{b}_3(t)$: 
\begin{equation}
    \tilde{q}_{3}(t)=\frac{\tilde{b}_3(t)+\tilde{b}_3^\dag(t)}{\sqrt{2}},
       ~~\tilde{p}_{3}(t)=\frac{\tilde{b}_3(t)-\tilde{b}_3^\dag(t)}{\sqrt{2}i}.
\notag
\end{equation}
The parameters and the initial conditions are set as follows: 
$\Delta=1$, $\lambda=0.01\Delta$, $\gamma=2.01\lambda$, and 
$c/L_4=0.1\Delta$. 
Also $\kappa$ is set to $\kappa=0.1\Delta$ (black lines in the figure) or 
$\kappa=0.01\Delta$ (light blue lines). 
The initial condition is $\mean{a_1(0)}=\mean{a_2^\dag(0)}=\mean{a_3^\dag(0)}
=\mean{a_4^\dag(0)}=1/\sqrt{2}$. 
Now the ideal oscillation \eqref{eq:a5 self oscillation} is represented 
in terms of the quadratures as 
\begin{equation}
    \left[
\begin{array}{c}
 \mean{\tilde{q}_3(t)} \\
 \mean{\tilde{p}_3(t)}
\end{array}
\right]=
\left[
\begin{array}{cc}
\cos{(\Delta t)} &\sin{(\Delta t)}  \\
-\sin{(\Delta t)} &\cos{(\Delta t )} 
\end{array}
\right]
\left[
\begin{array}{c}
 \mean{\tilde{q}_3(0)} \\
 \mean{\tilde{p}_3(0)}
\end{array}
\right].\notag
\end{equation}
With the initial values mentioned above, these are given by 
$\mean{\tilde{q}_3(t)}=\sqrt{\kappa}\cos{(\Delta t)}$ and $\mean{\tilde{p}_3(t)}
=-\sqrt{\kappa}\sin{(\Delta t)}$, meaning that there is a $\pi/2$ phase difference 
between the two quadratures. 
Figure \ref{fig:a5_self_oscillation} shows that the case $\kappa=0.01\Delta$ 
closely reproduces this ideal oscillation. 
In particular, the smaller value of $\kappa$ leads to the slower attenuation 
of the oscillation. 
This is simply because the smaller $\kappa$ is, the less amount of photons 
leaks out from the MCC with detuning $\Delta$. 
Thus, by setting $\kappa$ smaller, we can preserve the coherence of the 
light field oscillating with frequency $\Delta$ in the MCC. 
However, making $\kappa$ smaller also reduces the amount of photons coming 
from the NDPA into the MCC, and thus the amplitude of oscillation is limited. 
Conversely, a large value of $\kappa$ allows a flow of large amount of 
photons from NDPA to MCC, resulting in a large amplitude of the oscillation 
as demonstrated in the case $\kappa=0.1\Delta$ in Fig.~\ref{fig:a5_self_oscillation}. 
Therefore, there is a tradeoff between the coherence time and the amplitude 
of the self-oscillation.

\section{Functionalities 3: Active filters}
\label{sec.active filters}

As we have seen in Sec.~\ref{sec:Passive components}, the 2-input 2-output cavity 
works as a low-pass or high-pass filter with bandwidth $\kappa$ and maximal gain $1$. 
Here we show that, by the feedback amplification method, several types of filter with 
tunable bandwidth, gain, and phase, i.e., the quantum version of active filters, can 
be engineered.

\subsection{High-Q active filter}
\label{sec.high-Q filter}

First we show a simple first-order active filter. 
As in the quantum integrator, the controller is chosen as the high-pass filtering cavity 
\eqref{asymmetric HPF} with zero-detuning $\Delta=0$, which in this case is set to be asymmetric 
(i.e., $\kappa_1 \neq \kappa_2$). Then, the closed-loop system \eqref{main result} realized in the high-gain amplification limit 
is given by 
\begin{eqnarray}
& & \hspace*{-2.3em}
      \left[ \begin{array}{c}
                \tilde{b}_1(s) \\ 
                \tilde{b}_3^\dagger(s^\ast) \\
            \end{array} \right]
        = G^{\rm (fb)}(s)
            \left[ \begin{array}{c}
                b_1(s) \\ 
                b_4^\dagger(s^\ast) \\
            \end{array} \right], 
\nonumber \\ & & \hspace*{-2em}            
       G^{({\rm fb})}(s)       
        =  \frac{1}{s+(\kappa_2-\kappa_1)/2}
\nonumber \\ & & \hspace*{1em}
\label{active filter trans}
       \times 
            \left[ \begin{array}{cc}
                - s - (\kappa_1+\kappa_2)/2 & \sqrt{\kappa_1\kappa_2}  \\
                \sqrt{\kappa_1\kappa_2}    &  s - (\kappa_1 + \kappa_2)/2\\
            \end{array} \right].
\end{eqnarray}
Here we focus on the output $\tilde{b}^\dagger_3(s^\ast)$:  
\[
     \tilde{b}^\dagger_3(s^\ast) 
         = \frac{\sqrt{\kappa_1\kappa_2} }{s+(\kappa_2-\kappa_1)/2}b_1(s)
           + \frac{s-(\kappa_1+\kappa_2)/2 }{s+(\kappa_2-\kappa_1)/2}b^\dagger_4(s^\ast), 
\]
with $\kappa_2-\kappa_1>0$. 
Hence, this system functions as a low-pass filter for $b_1(s)$ with bandwidth 
$(\kappa_2-\kappa_1)/2$. 
In contrast to the standard low-pass filter \eqref{LPF} with bandwidth $\kappa$, 
the bandwidth of this active filter can be made very small by making $\kappa_1$ and 
$\kappa_2$ close to each other. 
As a result, the Q-factor can be largely enhanced from $Q=\omega_0/2\kappa$ to 
$Q'=\omega_0/(\kappa_2-\kappa_1)$. 
For instance for a coherent light field with frequency $\omega_0=3\times 10^{14}$ Hz, 
an optical cavity $\kappa=3\times 10^6$ leads to $Q=5\times 10^7$, 
while the active filter with $\kappa_1=\kappa$ and $\kappa_2=1.01\kappa$ leads to 
$Q'=1\times 10^{10}$. 
Note that this active filter also functions as an amplifier with gain 
$2\sqrt{\kappa_1\kappa_2}/(\kappa_2-\kappa_1)$, which becomes large if Q-factor 
increases. 
Importantly, in this case the idler noise mode $b_4$ is also amplified with gain 
$(\kappa_2+\kappa_1)/(\kappa_2-\kappa_1)$ at $s=0$. 
This means that basically the filtering makes sense only for an input field with amplitude 
much bigger than $(\kappa_2+\kappa_1)/(\kappa_2-\kappa_1)$. 
Also we remark that, as discussed in the case of integrator, the feedback loop now constructs 
another cavity, which should be taken into account for more precise modeling of the filter; 
we will give such a detailed investigation in 
Sec.~\ref{sec:Application to gravitational-wave detection} 
for another type of active filter discussed in the next subsection.


\subsection{Phase-cancellation filter}
\label{sec:Unstable filter}

The functionality provided by an active filter is not only modifying the gain profile, 
but changing the phase of an input filed. 
In fact Miao et al. proposed a quantum active filter that can effectively 
change the phase and thereby enhance the bandwidth of the gravitational 
wave detector or, in a wider sense, any cavity-based quantum sensor 
\cite{Miao 2015}. 
A rough description of their idea is as follows. 
When a gravitational wave propagates through the interferometer (arm cavity), 
then it must pick up a phase 
$\phi_\textrm{arm}(\Omega)=-2\Omega L_\textrm{arm}/c$, where $\Omega$, 
$L_\textrm{arm}$, and $c$ are the gravitational-wave frequency, the length of the cavity, 
and $c$ the speed of light, respectively. 
Note that only in Secs.~\ref{sec:Unstable filter} and \ref{sec:Application to gravitational-wave detection} 
we use the conventional $\Omega$ rather than $\omega$ to recall that 
this is the gravitational-wave frequency. 
This extra phase eventually limits the bandwidth of the detector; 
hence constructing an auxiliary intra-cavity filter with the transfer function 
$e^{-i\phi_\textrm{arm}(\Omega)}=e^{2i\Omega L_\textrm{arm}/c}$ will compensate this 
extra phase and thus may recover the bandwidth.

Here we show that the feedback amplification method can be employed to realize such a 
phase-cancelling filter in a fully optical setting. 
We again use the closed-loop system \eqref{active filter trans} and now consider 
the output $\tilde{b}_1$: 
\begin{align*}
     \tilde{b}_1(s)
         =& ~ G_{11}^\textrm{(fb)}(s)b_1(s)+G_{12}^\textrm{(fb)}(s)b_4^\dag(s^\ast)
\notag\\
         =& -\frac{ s + (\kappa_1+\kappa_2)/2 }{s+(\kappa_2-\kappa_1)/2}b_1(s)
           + \frac{\sqrt{\kappa_1\kappa_2} }{s+(\kappa_2-\kappa_1)/2}b^\dagger_4(s^\ast). 
\end{align*}
Let us then set $\kappa_2=0$:
\begin{equation}
\label{unstable filter}
     \tilde{b}_1(s) = 
     	G_{11}^\textrm{(fb)}(s)b_1(s) =
         -\frac{ s + \kappa_1/2 }{s -\kappa_1/2}b_1(s). 
\end{equation}
In the frequency domain $s=i\Omega$ this equation reduces to 
\[
        \tilde{b}_1(i\Omega)=G_{11}^\textrm{(fb)}(i\Omega)b_1(i\Omega)
            =-\frac{i\Omega+\kappa_1/2}{i\Omega-\kappa_1/2}b_1(i\Omega). 
\]
Then by setting $\Omega\ll\kappa_1$ and $\kappa_1=2c/L_\textrm{arm}$, we actually 
find that $G_{11}^\textrm{(fb)}$ approximates our target filter: 
\begin{align}
       G_{11}^\textrm{(fb)}(i\Omega)
              = &\frac{-\Omega^2+\kappa_1^2/4+i\Omega\kappa_1}
	                      {\Omega^2+\kappa_1^2/4}
            \approx\frac{\kappa_1^2/4+i\Omega\kappa_1}{\kappa_1^2/4}
\notag\\
         \approx & \, e^{4i\Omega/\kappa_1}
              = e^{2i\Omega L_\textrm{arm}/c}
              = e^{-i\phi_\textrm{arm}(\Omega)}.
\label{unstable filter transfer function}
\end{align}
This phase-cancelling filter might be realizable in practice by carefully devising the 
controller cavity so that the optical loss $\kappa_2$ is very small. 
Note that in the literature works \cite{Miao 2015,Shahriar 2018,Blair 2018,Miao 2019} 
an opto-mechanical oscillator was employed to realize the same filter where in that case 
$\kappa_2$ represents the magnitude of the thermal bath added on the oscillator; 
hence $\kappa_2 \approx 0$ requires the oscillator to be in an ultra-low temperature 
environment.

Lastly note that the system \eqref{unstable filter} is clearly unstable; particularly 
the system \eqref{unstable filter transfer function} represents a phase-lead filter that 
violates the causality. 
Similar to the case of integrator, therefore, in a practical setting such a phase-cancellation 
filter must be incorporated in a bigger system that is totally stable. 
In Sec.~\ref{sec:Application to gravitational-wave detection} we give a detailed study to see 
how much the phase-cancellation filter \eqref{unstable filter} could compensate the phase 
delay and thereby enhance the bandwidth of the stabilized gravitational-wave detector 
in a practical setup.


\subsection{Butterworth filter}

\begin{figure}[tb]
\centering\includegraphics[width=0.9\columnwidth]{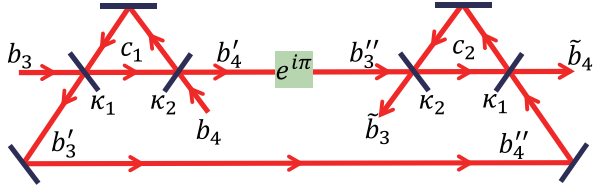}
\caption{
The controller $K$ for realizing the second order quantum Butterworth filter in the optical 
setting. 
The detuning of the left cavity is $\Delta$, while that of the right cavity is $-\Delta$. 
This controller has the inputs ($b_3$, $b_4$) and outputs ($\tilde{b}_3$, $\tilde{b}_4$). 
A phase shifter is embedded between the two cavities.}
\label{butterworthcontroller}
\end{figure}

Let us move back to the problem of modifying the gain profile via an active filter. 
A particularly useful bandpass filter, which is often used in classical electrical circuits, 
is the Butterworth filter. 
The transfer function of the $n$-th order classical Butterworth filter is given by 
$T_n(s)=g/B_n(s)$ where $g$ is a constant and the followings are examples of polynomials 
$B_n(s)$: 
\begin{align}
      B_1(s)=&s+1,\quad 
      B_2(s)=s^2+\sqrt{2}s+1,\notag\\
      B_3(s)=&(s+1)(s^2+s+1). \notag
\end{align}
The gain of the filter is given by 
\begin{equation}
     |T_n(i\omega)| = \frac{g}{\sqrt{(\omega/\omega_B)^{2n}+1}},\notag
\end{equation}
which has the steep roll-off characteristic of frequency, particularly for large $n$, at the 
cut-off frequency $\omega_B$.

A quantum version of Butterworth filter has actually been employed in the literature; 
in Ref.~\cite{Shapiro 2016}, a fourth-order quantum Butterworth filter was applied to 
enhance the channel capacity of a linear time-invariant bosonic channel. 
However, its physical realization has not been discussed. 
Here we show that, in the simple case $n=2$, the feedback amplification technique can be 
used to realize the quantum Butterworth filter.

The controller $K$ is chosen as the cascaded cavities, an optical case of which 
is depicted in Fig.~\ref{butterworthcontroller}. 
The left cavity with mode $c_1$ has two inputs ($b_3, b_4$) and two outputs ($b'_3, b'_4$), 
and the right one with mode $c_2$ has two inputs ($b''_3, b''_4$) and two outputs 
($\tilde{b}_3, \tilde{b}_4$). 
We assume that the detuning of the left and right cavities are $\Delta$ and $-\Delta$, 
respectively. 
A phase shifter $e^{i\pi}(=-1)$ is placed in the path from $b'_4$ to $b''_3$. 
Then the two input and output fields are connected as follows: 
\begin{equation}
    b''_3\mbox{}^\dag(s^\ast) = -b'_4\mbox{}^\dag(s^\ast), \quad 
     b''_4\mbox{}^\dag(s^\ast) = b'_3\mbox{}^\dag(s^\ast). \notag
\end{equation}
The input-output relations of each cavities are given by 
\begin{align}
       \left[\begin{array}{c}
           b'_3\mbox{}^\dag(s^\ast)  \\
           b'_4\mbox{}^\dag(s^\ast)
	\end{array}\right]
	&=K_l(s)
        \left[\begin{array}{c}
           b_3^\dag(s^\ast)  \\
           b_4^\dag(s^\ast)
	\end{array}\right],   \notag
\\
       \left[\begin{array}{c}
           \tilde{b}_3^\dag(s^\ast)  \\
           \tilde{b}_4^\dag(s^\ast)
       \end{array}\right]
       &=K_r(s)
       \left[\begin{array}{c}
          b''_3\mbox{}^\dag(s^\ast)  \\
          b''_4\mbox{}^\dag(s^\ast)
       \end{array}\right],  \notag
\end{align}
where $K_l(s)$ is given by the left hand side of Eq.~\eqref{asymmetric LPF} and
\begin{eqnarray}
& & \hspace*{-0.9em}
       K_r(s) =  \frac{1}{s+(\kappa_1+\kappa_2)/2+i\Delta}
\nonumber \\ & & \hspace*{1.2em}
        \times 
            \left[ \begin{array}{cc}
            	s + (\kappa_1 - \kappa_2)/2 + i\Delta & -\sqrt{\kappa_1\kappa_2} \\
                -\sqrt{\kappa_1\kappa_2} & s - (\kappa_1 - \kappa_2)/2 + i\Delta \\
            \end{array} \right].
\nonumber
\end{eqnarray}
The transfer function of the controller is thus given by 
\begin{equation}
    K(s)=K_r(s)
        \left[
	\begin{array}{cc}
	0 & -1  \\
	1 & 0
	\end{array}
	\right]
	K_l(s). \label{eq: controller for butterworth}
\end{equation}
Here we set $\Delta=(\kappa_1+\kappa_2)/2$. 
The feedback configuration is depicted in Fig.~\ref{quantum fb}, where $K(s)$ is now given 
by Eq.~\eqref{eq: controller for butterworth}. 
Then from Eqs.~\eqref{main result} and \eqref{eq: controller for butterworth}, we find that 
the output $\tilde{b}_3^\dag$ of the closed-loop system composed of this controller and 
a high-gain amplifier $G$ is given by
\begin{align}
     \tilde{b}_3^\dagger(s)
     =& ~ G_{21}^\textrm{(fb)}(s)b_1(s)+G_{22}^\textrm{(fb)}(s)b_4^\dag(s^\ast)
\notag\\
     =& -\frac{ \left\{ \kappa_1-\kappa_2+i(\kappa_1+\kappa_2) \right\}\sqrt{\kappa_1\kappa_2} }
                   {s^2 + (\kappa_2-\kappa_1)s + (\kappa_2-\kappa_1)^2/2}b_1(s)
\notag\\
       & - \frac{s^2 - (\kappa_2+\kappa_1)s + (\kappa_2+\kappa_1)^2/2}
                    {s^2 + (\kappa_2-\kappa_1)s + (\kappa_2-\kappa_1)^2/2}b^\dagger_4(s^\ast).
\notag
\end{align}
\begin{figure}[tb]
\centering\includegraphics[width=0.8\columnwidth]{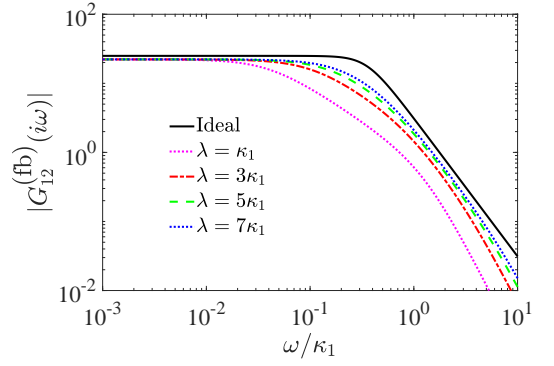}
\caption{
Gain plot of the ideal second-order quantum Butterworth filter (black solid line) and 
that of actual transfer function $G_{21}^\textrm{(fb)}=G_{21}K_{11}/(1-K_{21}G_{22})$ with 
several $\lambda$ (dotted lines). 
}
\label{butterworth_filter}
\end{figure}

The transfer function from $b_1$ to $\tilde{b}_3^\dag$ has a form of the second order 
Butterworth filter with cut-off frequency $\omega_B=(\kappa_2-\kappa_1)/\sqrt{2}$ and 
maximal gain $g=\sqrt{2\kappa_1\kappa_2(\kappa_1^2+\kappa_2^2)}/\omega_B^2$. 
Also, it is easy to see that the transfer function from $b_4^\dag$ to $\tilde{b}_1$ has the 
same form of second order Butterworth filter as above. 
Note that, as mentioned in Sec.~\ref{sec.high-Q filter}, the amplitude of the input field 
should be much bigger than that of the amplified idler vacuum noise.

Figure~\ref{butterworth_filter} shows the gain plot of the second-order Butterworth filter 
developed above. 
In this figure, the black solid line shows the gain plot of the ideal transfer function 
$G_{21}^\textrm{(fb)}(i\omega)=-K_{11}(i\omega)/K_{21}(i\omega)$, which corresponds to 
$\gamma \rightarrow 2\lambda+0$, while the dotted lines show the gain plot of 
$G_{21}^\textrm{(fb)}(i\omega)
=G_{21}(i\omega)K_{11}(i\omega)/\left\{1-K_{21}(i\omega)G_{22}(i\omega) \right\}$ 
with $\gamma=2.01\lambda$ and several parameters $\lambda=m\kappa_1$ ($m=1, 3, 5, 7$). 
The other parameters are fixed to $\kappa_2=1.5\kappa_1$ and 
$\Delta=(\kappa_1+\kappa_2)/2$. 
Now, as mentioned before, $|G_{11}(i\omega)|\gg1$ holds in the frequency range 
$\omega\ll\gamma=2.01\lambda$. 
Therefore, making $\lambda$ bigger results in broadening the frequency range where the 
approximation is valid, and in fact Fig.~\ref{butterworth_filter} shows that the dotted 
line approaches to the ideal solid line as $\lambda$ gets larger.


\subsection{Non-reciprocal amplifier}

\begin{figure}[t]
\includegraphics[width=0.87\columnwidth]{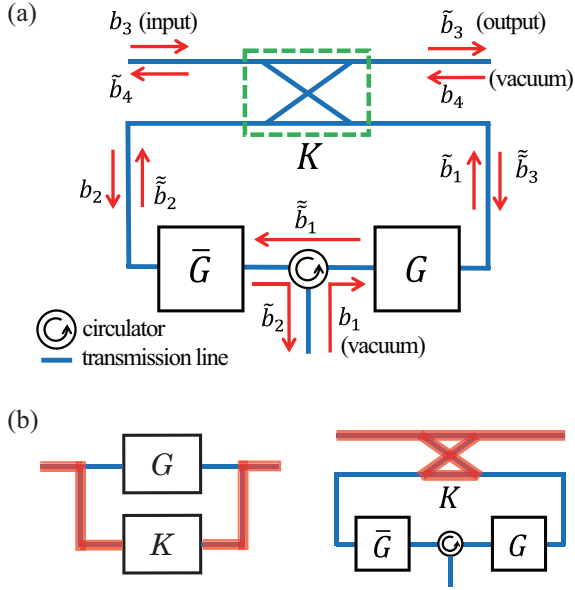}
\caption{
(a) Configuration of the proposed microwave non-reciprocal amplifier with the input $b_3$ and 
the amplified output $\tilde{b}_3$. 
This system is composed of two amplifiers ($G$, $\overline{G}$) and a passive controller $K$. 
Although in this figure $K$ represents a beam splitter, other general quantum passive systems 
can also work as $K$. 
(b) Intuitive illustration of the signal flow in the closed-loop system containing high-gain amplifiers 
in the general classical case (left) and in the proposed quantum case (right). 
}
\label{fig:a5_non_reciprocal_amp}
\end{figure}

The last topic in this section is a proposal of a non-reciprocal (directional) 
amplifier, constructed via the feedback amplification method. 
Non-reciprocal amplifiers are particularly important in the field of superconducting 
circuit based quantum technologies \cite{Kamal 2012,Lehnert 14,Korppi 2016}. 
In the microwave regime the phase-preserving amplifier obeys the same equation 
\eqref{input output relation of NDPA}, but the configuration differs from the optical 
case; the input $b_1$ and the corresponding output 
$\tilde{b}_1$ propagate along the same transmission line yet with opposite 
direction as shown in Fig.~\ref{integrator (a)(b)(c)(d)}(c). 
If the purpose of the use of amplifier is to detect a small signal $b_1$ generated from a 
source system, e.g., a superconducting qubit, then the propagating direction of the reflected 
output field $\tilde{b}_1$ must be changed to protect the source system from the backward 
field (if the output $\tilde{b}_1$ is the amplified vacuum noise field) or to measure it 
(if the output $\tilde{b}_1$ is the amplified signal). 
In fact there have been a number of theoretical and experimental proposals of 
active non-reciprocal amplifier \cite{Yurke, Abdo 14, Clerk 15,Malz 2018, Abdo 2018}. 
Our scheme is similar to \cite{Abdo 2018}, but it has a clear concept of using the 
feedback amplification to realize a robust non-reciprocal amplifier as described below.

Before describing the result, we note that the assumptions in the coherent feedback theory 
(e.g., see \cite{SLH review 2017}) are not always satisfied in superconducting devices, while 
there is certainly the case where the theory is valid and was experimentally demonstrated 
\cite{Kerckhoff 2012}. 
Here we leave this problem, i.e., the analysis for experimental implementability of the method 
in the proposed superconducting circuit, for the future work.

The proposed microwave non-reciprocal amplifier has a form of coherent feedback shown in 
Fig.~\ref{fig:a5_non_reciprocal_amp}(a), which takes into account the above-mentioned fact 
that the input and reflected output fields propagate along the same transmission line. 
This whole system has three inputs $(b_1, b_3, b_4)$ and three outputs 
$(\tilde{b}_2, \tilde{b}_3, \tilde{b}_4)$; 
particularly $b_3$ is the input signal and $\tilde{b}_3$ is the amplified signal to be detected, 
while $b_1$ and $b_4$ are the vacuum field. 
$G$ and $\overline{G}$ are both phase-preserving amplifiers, and $K$ is a passive system. 
As mentioned above, the source system generating $b_3$ may be contaminated due 
to the backward field $\tilde{b}_4$, which is not necessarily vacuum. 
Hence $\tilde{b}_4$ should be sufficiently suppressed.

Each system component has the following input-output relations:
\begin{align}
     \left[\begin{array}{c}
          \tilde{b}_1(s)\\
          \tilde{\tilde{b}}_1^\dag(s^\ast)
     \end{array}\right]=&
      \left[\begin{array}{cc}
          G_{11}(s) &G_{12}(s)  \\
          G_{21}(s) &G_{22}(s) 
      \end{array}\right]
      \left[\begin{array}{c}
          \tilde{\tilde{b}}_3(s) \\
          b_1^\dag(s^\ast)
       \end{array}\right],
\label{eq: G in non-reciprocal amp} \\
     \left[\begin{array}{c}
          \tilde{b}_2^\dag(s^\ast)\\
          \tilde{\tilde{b}}_2(s)
     \end{array}\right]=&
     \left[\begin{array}{cc}
          \overline{G}_{11}(s) &\overline{G}_{12}(s)  \\
          \overline{G}_{21}(s) &\overline{G}_{22}(s) 
     \end{array}\right]
     \left[\begin{array}{c}
          \tilde{\tilde{b}}_1^\dag(s^\ast) \\
          b_2(s)
      \end{array}\right],
\label{eq: Gbar in non-reciprocal amp} \\
      \left[\begin{array}{c}
          \tilde{b}_3(s) \\
          \tilde{\tilde{b}}_3(s)
       \end{array}\right]=&
       \left[\begin{array}{cc}
           K_{11}(s)&K_{12}(s)  \\
           K_{21}(s)&K_{22}(s) 
       \end{array}\right]
       \left[\begin{array}{c}
           b_3(s)  \\
       \tilde{\tilde{b}}_2(s)
       \end{array}\right],
\label{eq: K in non-reciprocal amp} \\
    \left[
        \begin{array}{c}
        \tilde{b}_4(s)  \\
         b_2(s)
        \end{array}
    \right]
    =&
    \left[
        \begin{array}{cc}
        K_{11}^\ast(s) & K_{21}^\ast(s) \\
        K_{12}^\ast(s) & K_{22}^\ast(s) 
        \end{array}
    \right]
    \left[
        \begin{array}{c}
        b_4(s)  \\
         \tilde{b}_1(s)
        \end{array}
    \right].
\label{eq: Kast in non-reciprocal amp}
\end{align}
Combining these equations, the input-output relation of the whole feedback 
controlled system is represented as follows; 
\begin{align}
          \left[ \begin{array}{c}
                \tilde{b}_2^\dagger(s^\ast) \\ 
                \tilde{b}_3(s) \\
                \tilde{b}_4(s) \\ 
            \end{array} \right]
        &= G^{\rm (fb)}(s)
            \left[ \begin{array}{c}
                b_1^\dag(s^\ast) \\ 
                b_3(s) \\
                b_4(s) \\ 
            \end{array} \right],  \notag\\
         G^\textrm{(fb)}(s)=&
            \left[\begin{array}{ccc}
               G_{11}^\textrm{(fb)}(s) &G_{12}^\textrm{(fb)}(s) &G_{13}^\textrm{(fb)}(s)  \\
               G_{21}^\textrm{(fb)}(s) &G_{22}^\textrm{(fb)}(s) & G_{23}^\textrm{(fb)}(s) \\
               G_{31}^\textrm{(fb)}(s) &G_{32}^\textrm{(fb)}(s) & G_{33}^\textrm{(fb)}(s)
            \end{array}\right].
        \label{eq: non-reciprocal amp before limit}
\end{align}
Here we found that, in a $s$-domain such that both $G(s)$ and $\overline{G}(s)$ have a large gain, 
the transfer function matrix $G^\textrm{(fb)}(s)$ converges to 
\begin{align}
\label{EQ:NON-RECIPROCAL AMP TF}
      &G^{\rm (fb)}(s)
      \notag \\
      & \hspace{2em} \to
        \left[
        \begin{array}{ccc}
            \dis -\frac{1}{K_{22}(s)} & \dis -\frac{K_{21}(s)}{K_{22}(s)} & 0  \\
            \dis -\frac{K_{12}(s)}{K_{22}(s)} & \dis \frac{\det{[K(s)]}}{K_{22}(s)} & 0 \\
            0 & 0 & \dis \frac{K_{11}^\ast(s) + \det{[K^\dag(s)]}}{1 + K_{22}^\ast(s)}
        \end{array}
        \right],        
\end{align}
where
\begin{equation}
    \Big|
        \frac{
            K_{11}^\ast(i\omega) + \det{[K^\dag(i\omega)]}
        }{
            1 + K_{22}^\ast(i\omega)
        }
    \Big|
    =
    1,
    \quad
    \forall \omega
    \label{EQ: G33 GAIN IS ONE}
\end{equation}
holds. 
The proof of Eq.~\eqref{EQ:NON-RECIPROCAL AMP TF} is given in 
Appendix \ref{sec:Proof of Eq non-reciprocal amp tf}, and Eq.~\eqref{EQ: G33 GAIN IS ONE} can be 
proven by using the unitary property of $K(i\omega)$ (i.e., 
$|K_{11}(i\omega)|^2 + |K_{12}(i\omega)|^2 = 1$, 
$|K_{21}(i\omega)|^2 + |K_{22}(i\omega)|^2 = 1$, and 
$K_{21}(i\omega)K_{11}^*(i\omega) + K_{22}(i\omega)K_{12}^*(i\omega)=0$). 
The point of this result is that, because $b_4$ is vacuum, the backward field mode $\tilde{b}_4$ 
propagating towards the input port (see Fig.~\ref{fig:a5_non_reciprocal_amp}(a)) is also a vacuum 
field in this high-gain limit. 
This means that the noise of the backward field at the input port is suppressed to the vacuum 
in the feedback configuration. 
Therefore, because the output signal $\tilde{b}_3$ contains the input signal $b_3$ with amplification 
gain $\det{[K]}/K_{22}$ that only depends on the passive component, this feedback-controlled 
system functions as a robust non-reciprocal amplifier or more broadly a robust non-reciprocal active 
filter.

Let us consider an example. 
If $K$ is given by a beam splitter with power transmissivity $T$: 
\begin{equation}
    K(s)=\left[\begin{array}{cc}
              \sqrt{T} & -\sqrt{1-T} \\
              \sqrt{1-T} &\sqrt{T} 
            \end{array}\right],
\notag
\end{equation}
then we have 
\begin{equation}
    G^\textrm{(fb)}(s)=
          \left[\begin{array}{ccc}
            -1/\sqrt{T} & -\sqrt{1/T-1} &0  \\
            \sqrt{1/T-1} & 1/\sqrt{T} & 0 \\
            0 &0 & 1
          \end{array}\right].\notag
\end{equation}
Hence, the input signal $b_3$ is amplified with amplification gain $1/\sqrt{T}$. 
Importantly, this non-reciprocal amplification is robust against the characteristic 
changes in the original amplifiers ($G$, $\overline{G}$) because the gain 
$1/\sqrt{T}$ is a tunable yet static quantity.

Lastly we provide a way for intuitively understanding the mechanism of non-reciprocity of the 
proposed amplifier. 
For this purpose, let us reconsider the classical feedback amplifier shown in Fig.~\ref{classical fb}. 
In this case, the input to the amplifier $G$, which is $u-Ky$, converges to zero when $G\to\infty$ 
due to the boundedness of the output $y$. 
This means that the signal injected to $G$ effectively vanishes in the high-gain limit; as a result, 
the signal flows only along the line illustrated by the orange thick line in the left figure of 
Fig.~\ref{fig:a5_non_reciprocal_amp}(b). 
This view implies that the proposed non-reciprocal amplifier has a similar characteristic. 
That is, the signal fields injected to the amplifiers $G$ and $\overline{G}$ effectively vanish (more 
precisely, suppressed to the vacuum field), and the signal propagates only through the passive system 
$K$ as illustrated by the orange thick line in the right figure of Fig.~\ref{fig:a5_non_reciprocal_amp}(b). 
This is indeed a useful property, because the signal does not pass through the circulator; 
in fact, microwave circulators are in general noisy, and thus, developing a non-reciprocal amplifier 
without circulators is what has been pursued recently 
\cite{Yurke, Abdo 14, Clerk 15,Malz 2018, Abdo 2018}. 
Therefore, it would be interesting to conduct an experiment to see how much the signals injected to the 
amplifiers $G$ and $\overline{G}$ are suppressed.


\section{Application to gravitational-wave detection}
\label{sec:Application to gravitational-wave detection}

As noted before, any functionality realized via the feedback amplification method 
should be evaluated in such a way that it is incorporated in a concrete setup 
with particular engineering purpose, to see its actual performance under practical 
constraints. 
Here we study the phase-cancellation filter discussed in Sec.~\ref{sec:Unstable filter}, 
and see how much it might broaden the bandwidth of the typical gravitational-wave 
detector.


\subsection{Basics of gravitational-wave detector}
\label{sec:Quantum noise in gravitational-wave detectors}

\begin{figure}[t]
\includegraphics[width=0.8\columnwidth]{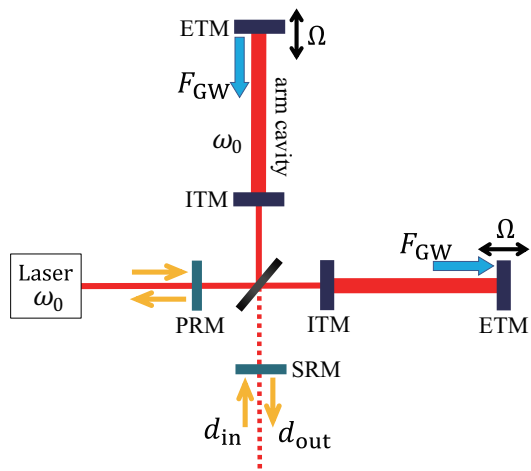}
\caption{
Schematic of the basic gravitational-wave detector. 
}
\label{LIGO}
\end{figure}

The most basic schematic of the gravitational-wave detector, particularly the one 
that uses a laser interferometer \cite{Abbott 2016, Harry 2010, Virgo 2015, Kagra 2012}, 
is shown in Fig.~\ref{LIGO}. 
The input laser with frequency $\omega_0$ is injected to the arm cavities through the 
power recycling mirror (PRM). 
Each arm cavity is composed of two mirrors: the input test mass (ITM) and the end 
test mass (ETM). 
A tidal force of gravitational-wave $F_\textrm{GW}$ with frequency $\Omega$ induces 
a pendulum motion of ETMs. 
Then the arm cavities create the signal light fields with frequency $\omega_0 + \Omega$, 
which are combined at the center half mirror and leak to outside through the signal 
recycling mirror (SRM); 
this output field is denoted as $d_\textrm{out}$. 
Note that a vacuum field $d_\textrm{in}$ unavoidably enters into the system at SRM.

\begin{figure}[tb]
\centering\includegraphics[width=0.64\columnwidth]{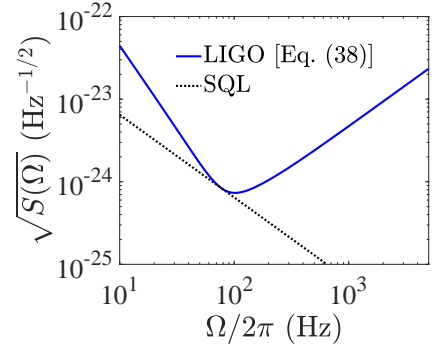}
\caption{
Quantum noise in the basic gravitational-wave detector (blue line). 
The black dashed line denotes SQL.}
\label{LIGO noise}
\end{figure}

The Hamiltonian of the entire system, in the rotating frame at frequency $\omega_0$, 
is given by \cite{Miao 2015, maphd} 
\begin{equation*}
	H=\frac{P^2}{2M}+\hbar\Delta_d d^\dag d
        -\hbar G_\textrm{arm}(d+d^\dag)X-F_\textrm{GW}X. 
\end{equation*}
$(X,P)$ are the differential (position, momentum) operators of ETMs, and they satisfy 
$[X(t),P(t)]=i\hbar$. 
$M$ is the mass of ETMs. 
$d$ is the sideband mode of the interferometer field, with detuning $\Delta_d$, 
which satisfies $[d(t),d^\dag(t)]=1$. 
$G_\textrm{arm}$ represents the coupling strength between $X$ and $d$, and it is given by 
$G_\textrm{arm}=\sqrt{2P_\textrm{arm}\omega_0/(\hbar cL_\textrm{arm})}$ with 
$P_\textrm{arm}$ the arm cavity power \cite{Miao 2015}. 
Then the dynamics of the system is given by
\begin{align}
     &  \dot{X}= \frac{1}{M}P, ~~
          \dot{P}= \hbar G_\textrm{arm}(d+d^\dag)+F_\textrm{GW},
\notag \\
     &  \dot{d}= -\left(i\Delta_d+\frac{\gamma_\textrm{IFO}}{2}\right)d
        	+iG_\textrm{arm}X-\sqrt{\gamma_\textrm{IFO}}d_\textrm{in},
\notag
\end{align}
where $\gamma_\textrm{IFO}$ is the coupling between $d$ and $d_\textrm{in}$. 
Also, the output equation of the system is given by 
\begin{equation}
      d_\textrm{out} = \sqrt{\gamma_\textrm{IFO}}d + d_\textrm{in}. \notag
\end{equation}
Note that 
$[d_\textrm{in}(t),d_\textrm{in}^\dag(t')]=[d_\textrm{out}(t),d_\textrm{out}^\dag(t')]
=\delta(t-t')$. 
The input-output relation in the Laplace domain, in terms of the quadratures 
$Q_d^\textrm{in,out}=(d_\textrm{in,out}+d_\textrm{in,out}^\dag)/\sqrt{2}$ and 
$P_d^\textrm{in,out}=(d_\textrm{in,out}-d_\textrm{in,out}^\dag)/\sqrt{2}i$, is obtained as 
\begin{equation*}
	\left[\begin{array}{c}
         Q_d^\textrm{out}(s)  \\
         P_d^\textrm{out}(s)
       \end{array}\right]
       =J(s)\left[\begin{array}{c}
           F_\textrm{GW}(s)  \\
           Q_d^\textrm{in}(s) \\
           P_d^\textrm{in}(s)
         \end{array}\right],
\end{equation*}
with
\begin{align*}
	&J(s)=\left[
	\begin{array}{ccc}
    	J_{11}(s) &J_{12}(s) &J_{13}(s)  \\
    	J_{21}(s) &J_{22}(s) &J_{23}(s) 
	\end{array}
	\right]
	=\notag\\
	&\left[
	\begin{array}{ccc}
	0 &\displaystyle \frac{s-\gamma_\textrm{IFO}/2}{s+\gamma_\textrm{IFO}/2} & 0 \\
	\displaystyle \frac{\sqrt{2\gamma_\textrm{IFO}}G_\textrm{arm}}{Ms^2(s+\gamma_\textrm{IFO}/2)} &\displaystyle \frac{-2\hbar G_\textrm{arm}^2\gamma_\textrm{IFO}}{Ms^2(s+\gamma_\textrm{IFO}/2)^2} & \displaystyle \frac{s-\gamma_\textrm{IFO}/2}{s+\gamma_\textrm{IFO}/2}
	\end{array}
	\right],
\end{align*}
where $\Delta_d=0$ is assumed. 
The gravitational-wave strain signal $h$, which is defined as 
$F_\textrm{GW}(t)=ML_\textrm{arm}\ddot{h}(t)$, can be detected by homodyne measuring 
$P_d^\textrm{out}$. 
The quantum noise operator is then defined as 
\begin{align}
     F_N(s)=&\frac{P_d^\textrm{out}(s)}{ML_\textrm{arm}s^2J_{21}(s)}-h(s)
\notag\\
               =& \, \Xi_Q(s)Q_d^\textrm{in}(s)+\Xi_P(s)P_d^\textrm{in}(s),
\label{def of Fn}
\end{align}
where
\begin{align}
       \Xi_Q(s)=-\frac{\sqrt{2\gamma_\textrm{IFO}}\hbar G_\textrm{arm}}
                                {ML_\textrm{arm}s^2(s+\gamma_\textrm{IFO}/2)}, 
\notag \\
       \Xi_P(s)=\frac{s-\gamma_\textrm{IFO}/2}
                              {\sqrt{2\gamma_\textrm{IFO}}G_\textrm{arm}L_\textrm{arm}}. 
\notag
\end{align}
Hence $F_N(s)$ is composed of the radiation pressure noise 
$\Xi_Q(s)Q_d^\textrm{in}(s)$ and the shot noise $\Xi_P(s)P_d^\textrm{in}(s)$, which are 
dominant in the low and high frequency range, respectively. 
The magnitude of $F_N(i\Omega)$ is quantified by the spectral density $S(\Omega)$, 
which is generally defined by \cite{Danilishin 2012,Aspelmeyer 2014,Wimmer 2014}
\begin{align}
      2\pi & S(\Omega)\delta(\Omega-\Omega')  
\notag\\
           =&\left\langle F_N(i\Omega)F_N^\dag(i\Omega') 
                  + F_N^\dag(i\Omega')F_N(i\Omega) \right\rangle/2. 
\label{noise definition}
\end{align}
It is now calculated as 
\begin{equation}
      S(\Omega)= \left( \left|\Xi_Q(i\Omega) \right|^2 + \left|\Xi_P(i\Omega) \right|^2 \right)/2,\label{eq:ligo noise proto}
\end{equation}
which is lower bounded by the standard quantum limit (SQL) 
\cite{maphd,haixingphd,Chen 2013}: 
$S_\textrm{SQL}(\Omega)=\left|\Xi_Q\Xi_P \right| = \hbar/(ML_\textrm{arm}^2\Omega^2)$. 
Note that, however, there have been several proposals to beat SQL, which can be now even 
experimentally observed in the real LIGO system \cite{LIGO 2020}. 
Figure~\ref{LIGO noise} shows the sensitivity $\sqrt{S(\Omega)}$ in the following typical setup 
\cite{Shahriar 2018,haixingphd}: $M=40\, \textrm{kg}$, $L_\textrm{arm}=4\, \textrm{km}$, 
$P_\textrm{arm}=800\, \textrm{kW}$, 
$\omega_0=2\pi c/\lambda_\textrm{laser}$, 
$\lambda_\textrm{laser}=1064\, \textrm{nm}$, 
$\Delta_d=0$, 
$\gamma_\textrm{IFO}=2\pi\times200\, \textrm{Hz}$. 
Also recall that $\Omega$ is the gravitational-wave frequency.

\subsection{Effect of the phase-cancellation filter}
\label{sec:Details of the unstable filter scheme}

As seen above, the detection sensitivity (roughly the inverse of the noise magnitude) is 
limited by the quantum noise. 
Especially, the following equality holds \cite{mizuno}, meaning that there is a tradeoff 
between the bandwidth and the peak sensitivity: 
\[
       \int_0^\infty \frac{1}{\left|\Xi_P(i\Omega) \right|^2 } d\Omega 
          = 2 \pi G_\textrm{arm}^2 L_\textrm{arm}^2. 
\]
In fact, because the integral does not depend on the bandwidth of the cavity, 
$\gamma_\textrm{IFO}$, a broad-band enhancement of the sensitivity is not allowed.

\begin{figure}[t]
\begin{minipage}{0.85\hsize}
\vspace{0.1cm}
\centerline{\normalsize(a)}
\includegraphics[width=\columnwidth]{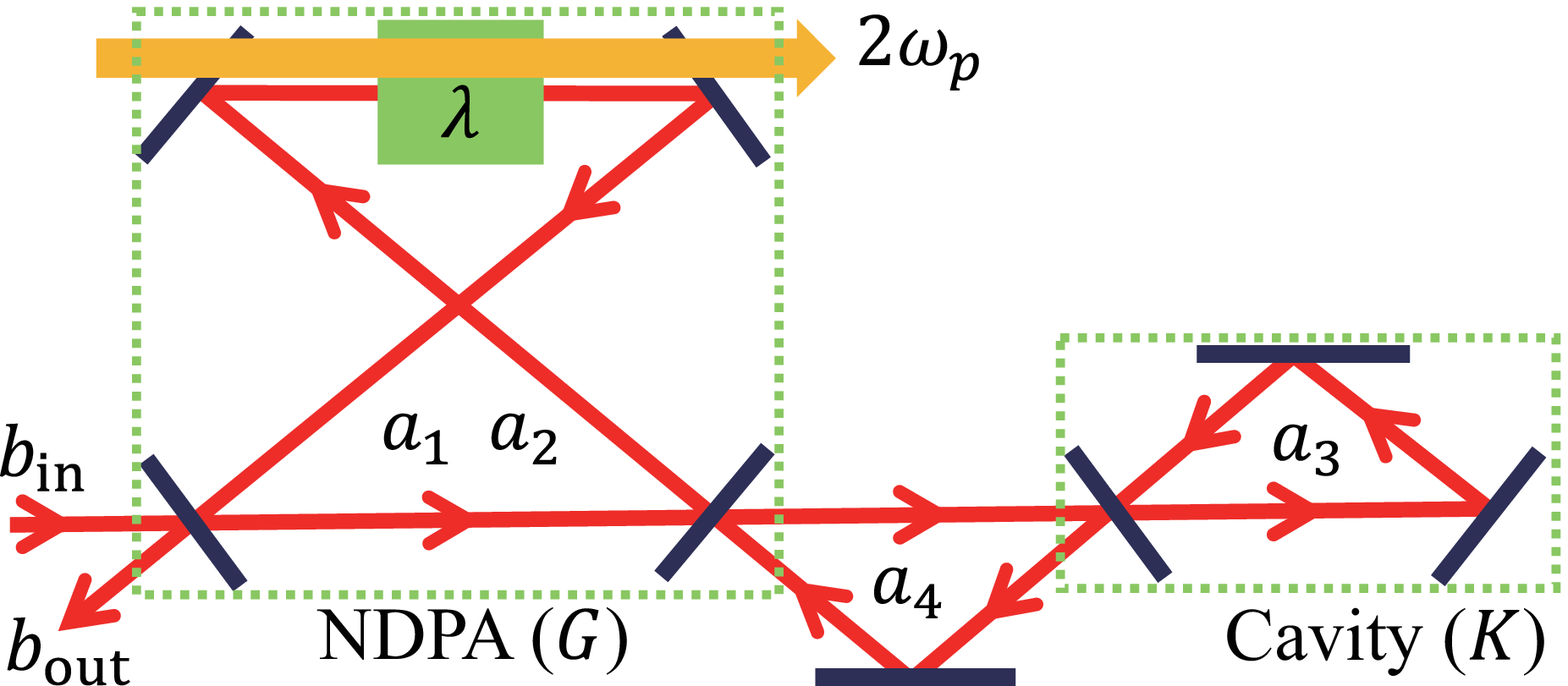}
\label{MCCFBNDPAcavity}
\end{minipage}

\begin{minipage}{\hsize}
\centerline{\normalsize(b)}
\includegraphics[width=\columnwidth]{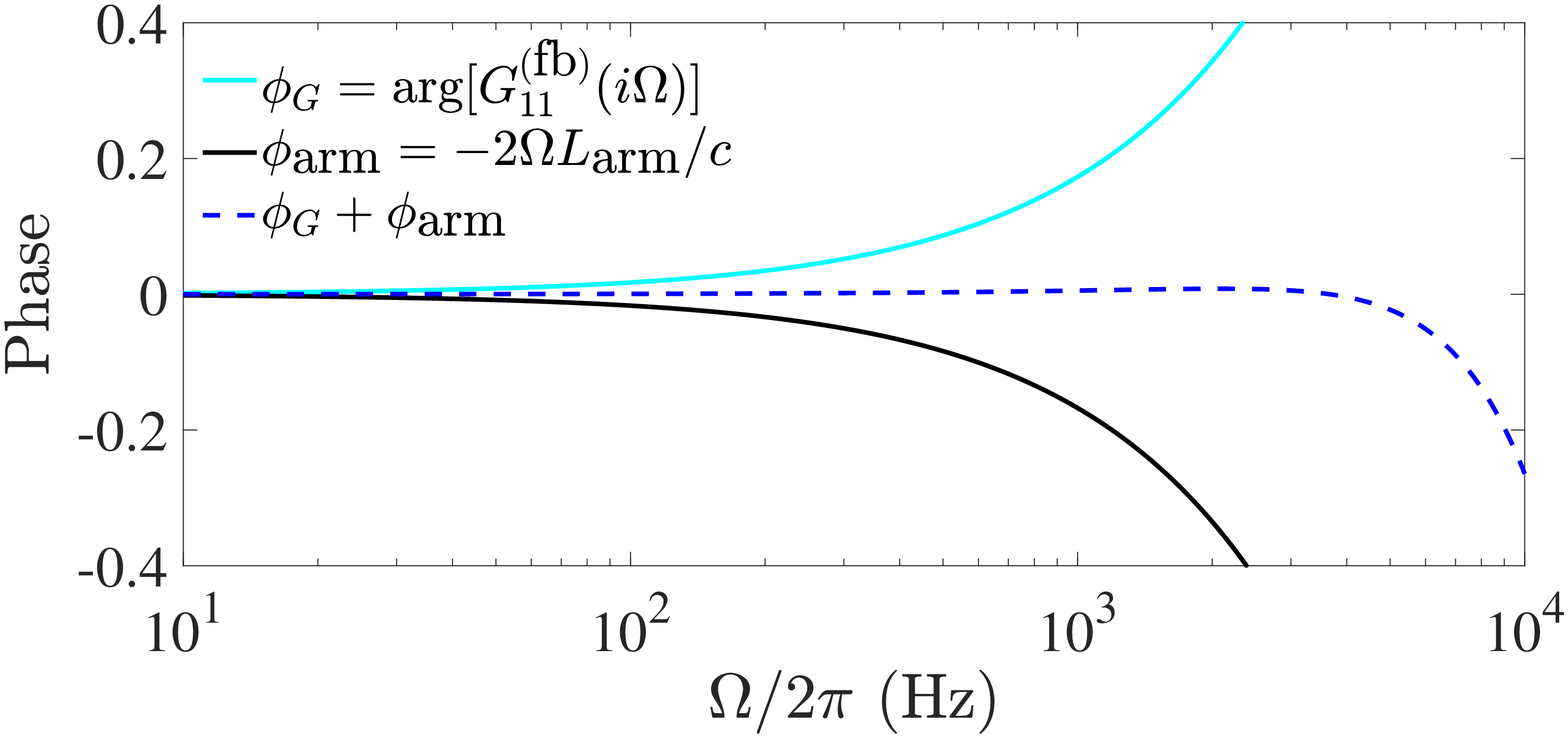}
\label{phase compensate}
\end{minipage}\\
\vspace{0.05em}
\begin{minipage}{\hsize}
\centerline{\normalsize(c)}
\includegraphics[width=\columnwidth]{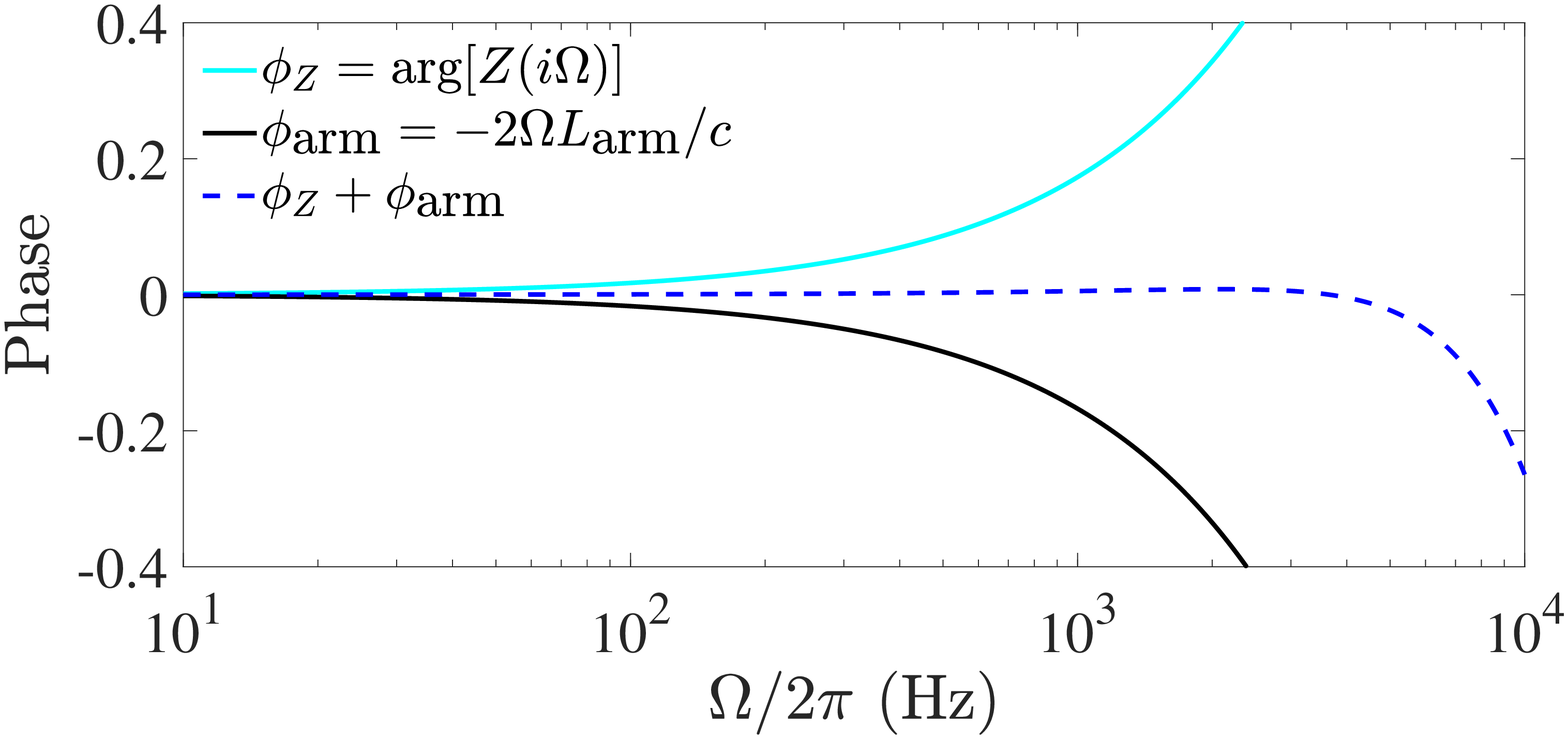}
\label{phace_cancel_cavity_matlab}
\end{minipage}
\caption{
(a) Configuration of the phase-cancellation filter, where the feedback loop between the amplifier 
and the cavity is regarded as another cavity with mode $a_4$. 
(b) Phase plot of the phase-cancellation filter $G_{11}^\textrm{(fb)}(i\Omega)$ in Eq.~\eqref{phi G} 
together with $\phi_\textrm{arm}(\Omega)$. 
(c) Phase of the phase-cancellation filter $Z(i\Omega)$ in 
Eq.~\eqref{MCCFBNDPAcavity laplace domain input output relation} 
together with $\phi_\textrm{arm}(\Omega)$. 
}
\label{unstable filter (a) to (d)}
\end{figure}

As described in Sec.~\ref{sec:Unstable filter}, the above tradeoff is attributed to the 
frequency-dependent propagation phase 
$\phi_\textrm{arm}(\Omega)=-2\Omega L_\textrm{arm}/c$. 
The idea proposed in Ref.~\cite{Miao 2015} is to construct the phase-cancellation filter 
with transfer function $e^{-i\phi_\textrm{arm}(\Omega)}=e^{2i\Omega L_\textrm{arm}/c}$ 
to compensate $\phi_\textrm{arm}(\Omega)$. 
Now, unlike the optomechanics-based scheme proposed in Ref.~\cite{Miao 2015}, we can 
construct the same filter \eqref{unstable filter transfer function} in all-optics setup, using 
the feedback amplification method. 
Figure \ref{unstable filter (a) to (d)}(b) shows $\phi_\textrm{arm}(\Omega)$ and 
\begin{equation}
\label{phi G}
        \phi_G
          = \textrm{arg}\left[G_{11}^\textrm{(fb)}(i\Omega)\right]
          = \textrm{arg}
               \left[\frac{ G_{11} - K_{21} {\rm det}[G]}{1-K_{21}G_{22}}\right], 
\end{equation}
where $G$ is the transfer function of NDPA and $K$ is given by 
Eq.~\eqref{asymmetric HPF} with $\Delta=\kappa_2=0$; 
this approximates $-\phi_\textrm{arm}$ in the high-gain limit, as proven in 
Eq.~\eqref{unstable filter transfer function}. 
The parameters are set as 
$L_\textrm{arm}=4$ km, $\lambda=3\times10^6$ Hz, $\gamma=2.01\lambda$, and 
$\kappa_1=2c/L_\textrm{arm}$. 
We can see from Fig.~\ref{unstable filter (a) to (d)}(b) that the filter certainly 
achieves the desired phase cancellation in the frequency range where 
$\Omega\ll\kappa_1=2c/L_\textrm{arm}\approx2\pi\times 2.39\times10^4$ Hz 
is satisfied.

Now recall that the filter is realized as the feedback-controlled system shown in 
Fig.~\ref{unstable filter (a) to (d)}(a). 
That is, as discussed in the case of integrator in Sec.~\ref{sec:Integrator}, the feedback 
loop between the amplifier ($G$) and the cavity ($K$) forms a loop cavity with mode $a_4$. 
The total Hamiltonian of the filter is given by Eq.~\eqref{three cascaded cavities}, and 
we again assume $\omega_k=\omega_p$ ($k=$1, \ldots, 4). 
Then in the rotating frame at frequency $\omega_p$, the dynamics and output equation of the filter are given by
\begin{align}
    & \dot{a}_1= -\frac{\gamma}{2}a_1+\lambda a_2^\dag-\sqrt{\gamma}b_\textrm{in}, ~~
        \dot{a}_2^\dag = \lambda a_1+ig_{24}a_4^\dag,
\notag \\
    & \dot{a}_3^\dag = ig_{34}a_4^\dag, ~~
        \dot{a}_4^\dag = ig_{24}a_2^\dag+ig_{34}a_3^\dag,
\notag \\
     & b_\textrm{out} = \sqrt{\gamma}a_1 + b_\textrm{in}, 
\notag
\end{align}
with $g_{24}=\sqrt{c\gamma/L_4}$ and $g_{34}=\sqrt{c\kappa_1/L_4}$. 
This equation is the same as Eq.~\eqref{three cavities dynamics} except that control cavity $K$ 
couples to only the loop cavity. 
The input-output relation of this system in the Laplace domain is represented as
\begin{align}
       b_\textrm{out}(s) &= Z(s) b_\textrm{in}(s), 
\notag \\
      Z(s) & =\frac{s^4+\alpha_3s^3+\alpha_2s^2+\alpha_1s+\alpha_0}
                      {s^4+\beta_3s^3+\beta_2s^2+\beta_1s+\beta_0},
\label{MCCFBNDPAcavity laplace domain input output relation}
\end{align}
where
\begin{align}
    \alpha_3=& -\gamma/2, ~
    \alpha_2=g_{24}^2+g_{34}^2-\lambda^2, ~ 
    \alpha_1=-(g_{24}^2+g_{34}^2)\gamma/2, 
\notag\\
     \alpha_0=&-\lambda^2g_{34}^2, ~
     \beta_3=\gamma/2, ~ 
     \beta_2=g_{24}^2+g_{34}^2-\lambda^2, 
\notag\\
     \beta_1=& (g_{24}^2+g_{34}^2)\gamma/2, ~
     \beta_0=-\lambda^2g_{34}^2.
\notag
\end{align}
Now we show that $Z(i\Omega)$ approximates the target phase-cancellation filter 
$e^{2i\Omega L_\textrm{arm}/c}$ in the high-gain limit $\gamma\to2\lambda+0$. 
First, by setting $\kappa_1=2c/L_\textrm{arm}$ and taking this limit, the coefficients 
in $Z(s)$ become 
\begin{align}
      \alpha_3 = &-\beta_3=-\lambda, ~
      \alpha_2 = \beta_2=\frac{2c(c+L_\textrm{arm}\lambda)}
                                      {L_\textrm{arm}L_4}-\lambda^2,
\notag \\
       \alpha_1= &-\beta_1=-\frac{2c\lambda(c+L_\textrm{arm}\lambda)}
                                           {L_\textrm{arm}L_4}, ~
     \alpha_0 = \beta_0=-\frac{2c^2\lambda^2}{L_\textrm{arm}L_4}. 
\notag
\end{align}
Then, in the $s$-domain with $|s|\ll\gamma$, or equivalently $|s|\ll\lambda$, we have 
\[
      Z(s) \approx \frac{\alpha_1s+\alpha_0}{\beta_1s+\beta_0}
         =-\frac{\displaystyle s+\frac{c\lambda}{c+L_\textrm{arm}\lambda}}
                    {\displaystyle s-\frac{c\lambda}{c+L_\textrm{arm}\lambda}}.
\]
Now we set an additional assumption $\kappa_1\ll\gamma$, which leads to 
$c\ll L_\textrm{arm}\lambda$ and as a result 
\begin{equation}
        Z(s)\approx-\frac{\displaystyle s+c/L_\textrm{arm}}
                                  {\displaystyle s-c/L_\textrm{arm}}.\notag
\end{equation}
This is exactly the same as the transfer function $G_\textrm{11}^{\textrm{(fb)}}(s)$ in 
Eq.~\eqref{unstable filter} with $\kappa_1=2c/L_\textrm{arm}$. 
Hence $Z(i\Omega)\approx e^{2i\Omega L_\textrm{arm}/c}=e^{-i\phi_\textrm{arm}(\Omega)}$ 
holds, meaning that the system depicted in Fig.~\ref{unstable filter (a) to (d)}(a) may 
approximate the target phase-cancellation filter. 
This can be actually seen in Fig.~\ref{unstable filter (a) to (d)}(c) showing the phase 
plot of $Z(i\Omega)$ given in Eq.~\eqref{MCCFBNDPAcavity laplace domain input output relation}, 
where $L_4=0.5$ m and the other parameters are the same as those used in 
Fig.~\ref{unstable filter (a) to (d)}(b). 
This shows that the exact model incorporating the loop cavity $a_4$ certainly 
has the desired phase cancelling effect.


\subsection{The entire system and stabilizing control}
\label{sec:The entire system and stabilizing control}

\begin{figure}[tb]
\centering\includegraphics[width=\columnwidth]{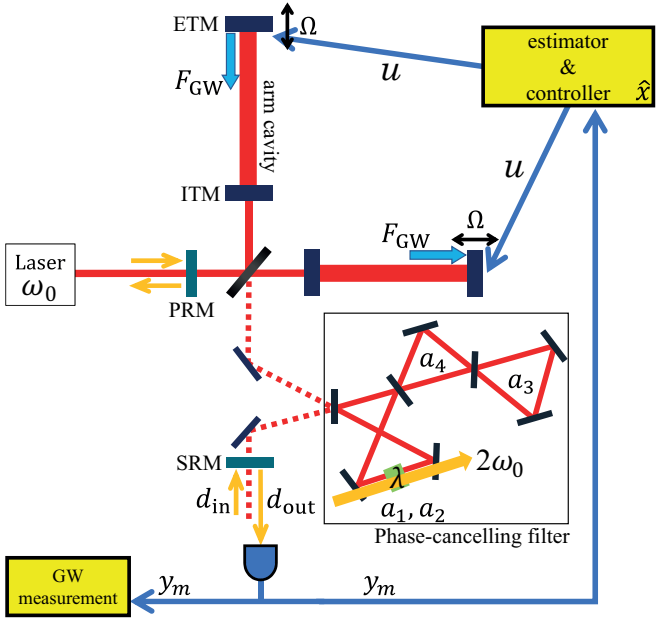}
\caption{Structure of the entire controlled system. }
\label{LIGONDPAcontrol}
\end{figure}

In the previous subsection we have seen that the constructed filter certainly has a desired 
phase-cancellation property, from which we expect that this active filter broadly enhances 
the sensitivity of the gravitational-wave detector in the high-frequency regime. 
Here we model the entire system composed of the interferometer and the phase-cancellation 
filter depicted in Fig.~\ref{LIGONDPAcontrol}. 
This entire system must be stabilized, because the phase cancellation filter is itself an unstable 
system; 
here we employ measurement-based feedback for this purpose. 

Note that this measurement feedback is not an additional requirement over the existing proposals; 
the stabilization is necessary as well in the optomechanics-based implementation 
\cite{Miao 2015, Shahriar 2018}. 

Let us begin with the dynamics of the entire system without stabilization. 
Here we assume that $\omega_p=\omega_k=\omega_0$ ($k=$1, \ldots, 4). 
Then in the rotating frame at frequency $\omega_0$, the Hamiltonian of the entire system 
is given by 
\begin{align*}
       H_\textrm{tot}=&\frac{P^2}{2M}+\Delta_d d^\dag d
                -\hbar G_{\textrm{arm}}(d+d^\dagger)X-F_\textrm{GW}X
\notag\\
        & + \hbar g_\textrm{NI}(d^\dagger a_1+da_1^\dagger)
           +i \hbar\lambda(a_1^\dagger a_2^\dagger-a_1a_2)
\notag\\
        & + \hbar g_{24}(a_2^\dagger a_4+a_2a_4^\dagger)
            + \hbar g_{34}(a_3^\dagger a_4+a_3a_4^\dagger), 
\end{align*}
where again $(X, P)$ are the differential (position, momentum) operators of ETMs and 
$d$ is the sideband mode of the interferometer field. 
We assume that only $a_1$ couples with $d$, with strength 
$g_\textrm{NI}=\sqrt{c\gamma/(2L_\textrm{arm})}$. 
The signal leaks to outside through the SRM where the vacuum input $d_\textrm{in}$ 
must enter. 
Then the dynamical equation of the entire system is given by
\begin{align*}
&       \dot{X}= \frac{1}{M}P, ~
       \dot{P}= \hbar G_{\textrm{arm}}(d+d^\dagger)+F_\textrm{GW},  
\\
&       \dot{d}=  -i\Delta_d d-\frac{\gamma_{\textrm{IFO}}}{2}d+iG_{\textrm{arm}}X
                        -ig_\textrm{NI}a_1-\sqrt{\gamma_{\textrm{IFO}}}d_{\textrm{in}},
\\
&       \dot{a}_1= -\frac{\gamma_\textrm{1loss}}{2}a_1-ig_\textrm{NI}d
                         +\lambda a_2^\dagger-\sqrt{\gamma_\textrm{1loss}}b_\textrm{1loss},
\\
&        \dot{a}_2 = \lambda a_1^\dagger-ig_{24}a_4, ~ 
        \dot{a}_3 = -\frac{\kappa_\textrm{3loss}}{2}a_3-ig_{34}a_4
                         -\sqrt{\kappa_\textrm{3loss}}b_\textrm{3loss},
\\
&        \dot{a}_4 = -\frac{\kappa_\textrm{4loss}}{2}a_4
                          -ig_{24}a_2-ig_{34}a_3-\sqrt{\kappa_\textrm{4loss}}b_\textrm{4loss},
\\
&        d_\textrm{out} = \sqrt{\gamma_\textrm{IFO}}d + d_\textrm{in}, 
\end{align*}
where $b_{k\textrm{loss}}$ ($k=$1, 3, 4) are the noise field representing the optical losses 
of the internal modes $a_k$ with magnitude $\kappa_{k\textrm{loss}}$. 
We use the quadrature representation 
$q_k=(a_k+a_k^\dag)/\sqrt{2}$, $p_k=(a_k-a_k^\dag)/(\sqrt{2}i)$ ($k=d, 1, 2, 3, 4$), 
$Q_d^\textrm{in,out}=(d_\textrm{in,out}+d_\textrm{in,out}^\dag)/\sqrt{2}$, 
$P_d^\textrm{in,out}=(d_\textrm{in,out}-d_\textrm{in,out}^\dag)/(\sqrt{2}i)$, 
$Q_{n\textrm{loss}}=(b_{n\textrm{loss}}+b_{n\textrm{loss}}^\dag)/\sqrt{2}$, 
$P_{n\textrm{loss}}=(b_{n\textrm{loss}}-b_{n\textrm{loss}}^\dag)/(\sqrt{2}i)$ ($n=1, 3, 4$). 
Also we define the dimensionless operators $X_M=X\sqrt{M\Omega_M/\hbar}$ and 
$P_M=P/\sqrt{\hbar M\Omega_M}$, with $\Omega_M$ the eigenfrequency of the ETM; 
they satisfy $[X_M,P_M]=i$. 
Then the above dynamical equations are summarized to 
\begin{equation}
        \dot{x}=Ax+B_ww,\quad y=Cx+Dw,\notag
\label{LIGONDPA dynamics}
\end{equation}
where $x=[\; X_M\; P_M\; q_d\; p_d\; q_1\; p_1\; q_2\; p_2\; q_3\; p_3\; q_4\; p_4\; ]^T$, 
$w=[\; F_\textrm{GW}\; Q_d^\textrm{in}\; P_d^\textrm{in}\; Q_{1\textrm{loss}}\; P_{1\textrm{loss}}\; 
Q_{3\textrm{loss}}\; P_{3\textrm{loss}}\; Q_{4\textrm{loss}}\; P_{4\textrm{loss}}\; ]^T$, and
$y=[\; Q_d^\textrm{out}\; P_d^\textrm{out}\; ]^T$. 
The matrices $A\in\mathbb{R}^{12\times12}$, $B_w\in\mathbb{R}^{12\times9}$, 
$C\in\mathbb{R}^{2\times12}$, $D\in\mathbb{R}^{2\times9}$ are shown in 
Appendix~\ref{sec:The matrix entries of ABCD}. 
Note that $A$ has eigenvalues with positive real part, meaning that the uncontrolled 
entire system is unstable.

To stabilize the system, we apply the measurement-based quantum feedback control, 
particularly the quantum linear quadratic Gaussian (LQG) feedback control 
\cite{Nurdin NY book}, which has the same form as the classical version 
\cite{Athans 1971}. 
This control is generally conducted by feeding a measurement output back to control the 
system. 
In our case we measure $P_d^\textrm{out}$ by the photodetector 
(note that measuring both $Q_d^\textrm{out}$ and $P_d^\textrm{out}$ is prohibited by quantum mechanics); 
the measurement result is used to construct the estimate $\hat{x}$, which is fed back to 
control the ETMs directly by implementing a piezo-actuator \cite{ETM control}. 
This control is modeled by adding the classical input $u=-F_u\hat{x}$ to the dynamics 
of the oscillator, where $F_u\in\mathbb{R}^{1\times12}$ is the feedback gain to be designed. 
In the LQG setting, the (quantum) Kalman filter is used to obtain the least squared estimate 
$\hat x$. 
The entire controlled system are then given by
\begin{align}
      & \dot{x} = Ax+B_ww+B_uu, \label{eq: state space gw}
      \\
      &  y_m =  P_d^\textrm{out}=C_mx+D_mw,
\notag \\
      & \dot{\hat{x}} = A\hat{x}+B_u u+K_u(y_m-C_m\hat{x}), ~~ u=-F_u\hat{x}, 
\notag
\end{align}
where $K_u\in\mathbb{R}^{12}$ is the Kalman gain shown later. 
$B_u=[\,0, 1, 0, \cdots, 0]^T\in\mathbb{R}^{12}$ (only the second element is 
non-zero) represents that the actuator directly drives $P_M$ of the oscillator. 
$C_m\in\mathbb{R}^{1\times12}$ and $D_m\in\mathbb{R}^{1\times9}$ are second 
row vectors of $C$ and $D$, respectively. Here we define $e=\hat{x}-x$. 
Then the above dynamical equation is rewritten as 
\begin{equation*}
	\left[
	\begin{array}{c}
	\dot{x}  \\
	 \dot{e}
	\end{array}
	\right]=A_\textrm{tot}
	\left[
	\begin{array}{c}
	x  \\
	 e
	\end{array}
	\right]+B_\textrm{tot}w, ~~
	y_m=C_\textrm{tot}
	\left[
	\begin{array}{c}
	x  \\
	 e
	\end{array}
	\right]+D_\textrm{tot}w,
\end{equation*}
where
\begin{align}
	A_\textrm{tot}=&\left[
	\begin{array}{cc}
	A-B_uF_u &-B_uF_u  \\
	0 &A-K_uC_m
	\end{array}
	\right], ~~
	C_\textrm{tot}=\left[
	\begin{array}{cc}
	C_m &0 
	\end{array}
	\right],
\notag \\ 
	B_\textrm{tot}=&\left[
	\begin{array}{c}
	B_w  \\
	 K_uD_m-B_w
	\end{array}
	\right], ~~
       D_\textrm{tot}=D_m.
\notag
\end{align}
The entire system becomes stable when $A_\textrm{tot}$ has no eigenvalue with 
positive real part. 
Since the eigenvalues of $A_\textrm{tot}$ are the same as those of $A-B_uF_u$ and 
$A-K_uC_m$, we can stabilize the system by determining appropriate $F_u$ and $K_u$. 
The necessary and sufficient condition for such $F_u$ and $K_u$ to exist is that 
the system is controllable and observable; that is, the following controllability 
matrix $\mathcal{C}_u$ and observability matrix $\mathcal{O}_{y_m}$ are of full-rank: 
\begin{align}
   & \mathcal{C}_u
         =\left[\begin{array}{ccccc}
               B_u & AB_u & \cdots  & A^{11}B_u 
           \end{array}\right], 
\notag \\
   & \mathcal{O}_{y_m}
         =\left[\begin{array}{ccccc}
               C_m^T & A^T C_m^T & \cdots  & (A^T)^{11}C_m^T 
           \end{array}\right]^T. 
\label{Controllability and observability matrices}
\end{align}
In the LQG setup, $F_u$ and $K_u$ are determined from the policy to minimize the 
following cost function $\mathcal{J}$ and the estimation error $\epsilon$: 
\begin{align}
       \mathcal{J} = &  \lim_{t\to \infty} \frac{1}{t}
            \Big\langle \int_0^t \left(x^T(\tau)Qx(\tau)+Ru^2(\tau)\right)d\tau \Big\rangle, 
\notag \\
        \epsilon = & \left\langle (x-\hat{x})^T(x-\hat{x})\right\rangle, \label{eq: estimate error}
\end{align}
where $Q\in\mathbb{R}^{12\times12}$ and $R\in\mathbb{R}$ are the weighing matrices. 
From the separation principle of the LQG control, these two optimization problems 
can be solved separately. 
If the optimal solutions of $F_u$ and $K_u$ are uniquely determined, then they stabilize 
the entire system and are given by 
\begin{equation*}
       F_u=R^{-1}B_u^TP_F, ~~
       K_u=(P_KC_m^T+B_wVD_m^T)(D_mD_m^T)^{-1},
\end{equation*}
where $P_F\in\mathbb{R}^{12\times12}$ and $P_K\in\mathbb{R}^{12\times12}$ are the 
solutions of the following algebraic Riccati equations:
\begin{align}
      &P_FA+A^TP_F-P_FB_uR^{-1}B_u^TP_F+Q=0, 
\notag \\
      &P_KA^T+AP_K+B_wVB_w^T-(P_KC_m^T+B_wVD_m^T)
\notag\\
      & ~~~~ \times(D_mVD_m^T)^{-1}(P_KC_m^T+B_wVD_m^T)^T=0.
\notag
\end{align}
$V$ is the covariance matrix of the vector $w$. 
Note that $F_\textrm{GW}$, the first element of $w$, is assumed to be a Gaussian noise with 
known variance, but in reality it is an unknown signal whose noise part is not necessarily Gaussian; 
as we will describe later, this assumption can be weakened so that only a stabilizing controller exists.


\subsection{Quantum noise of the stabilized system}
\label{sec:Quantum noise of the stabilized system}

\begin{figure}[tb]
\centering
\includegraphics[width=0.7\columnwidth]{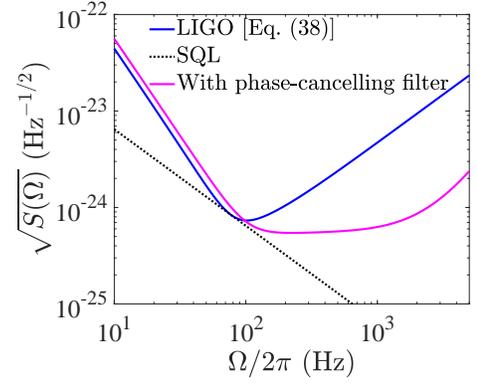}
\caption{
Quantum noise spectral density of the controlled gravitational-wave detector, containing 
the phase-cancellation filter. 
}\label{LIGONDPAnoise_with_loss}
\end{figure}

The quantum noise observed at the detector is calculated as follows. 
First we have 
\begin{align}
        y_m(s)=&\left[C_\textrm{tot}(sI-A_\textrm{tot})^{-1}B_\textrm{tot}
                +D_\textrm{tot}\right]w(s)
\notag\\
          = & \, \Psi_{Q_d}(s)Q_d^\textrm{in}(s)+\Psi_{P_d}(s)P_d^\textrm{in}(s) + \Psi_h(s)h(s) 
\notag\\
          & + \sum_{k=1,3,4} \Big( 
             \Psi_{Q_k}(s)Q_{k\textrm{loss}}(s) + \Psi_{P_k}(s)P_{k\textrm{loss}}(s) \Big),
\notag
\end{align}
where the functions $\Psi_{\star}$ are the transfer functions from the 
corresponding noise fields and the gravitational-wave strain signal $h(t)$ 
to the output $y_m$. 
As shown in Eq.~\eqref{def of Fn}, the quantum noise operator is defined as 
$F_N(i\Omega)=y_m(i\Omega)/\Psi_h(i\Omega)-h(i\Omega)$. 
Then from Eq.~\eqref{noise definition}, we obtain the noise spectral density:
\begin{align}
        S(\Omega)=& \frac{1}{2|\Psi_h|^2} \Big(|\Psi_{Q_d}|^2+|\Psi_{P_d}|^2+|\Psi_{Q_1}|^2+|\Psi_{P_1}|^2
\notag\\
            &\quad\quad\quad + |\Psi_{Q_3}|^2+|\Psi_{P_3}|^2+|\Psi_{Q_4}|^2
               +|\Psi_{P_4}|^2 \Big).
\notag
\end{align}

The value of parameters chosen in this study are shown in Table~\ref{tab: parameters setup}. 
For the interferometer part, unlike the setup in Fig.~\ref{LIGO noise}, a non-zero value of $\Delta_d$ 
is taken, which is necessary for the system to be controllable and observable; 
actually ${\mathcal C}_u$ and ${\mathcal O}_{y_m}$ in 
Eq.~\eqref{Controllability and observability matrices} are both of full-rank in these parameter choice. 
This non-zero value of $\Delta_d$ and the value of $\gamma_\textrm{IFO}$ are calculated from the 
scaling law \cite{scaling law, haixingphd} of the gravitational-wave detector containing a SRM; 
consequently the coupling constant $\gamma_\textrm{IFO}$ is effectively changed to 1062 Hz from 
$2\pi\times 200$ Hz. 
For the phase-cancellation filter part, we emphasize that $\kappa_1 = 2c/L_\textrm{arm}$ is the 
condition to cancel the phase and it draws the connection between the interferometer and the 
phase-cancellation filter. 
For the LQG controller part, the value of $1/2$ in $V={\rm diag}\{10^{-22}, 1/2, \cdots, 1/2\}$ 
(all $1/2$ except the $(1,1)$ element) denotes the vacuum fluctuation. 
On the other hand, the $(1,1)$ element of $V$ denotes the variance of $F_\textrm{GW}$, which 
is unknown as mentioned at the end of the previous subsection. 
Hence we used Fig.~1 in Ref.~\cite{Abbott 2016} to have an estimate value $10^{-22}$, meaning 
that the Kalman filter does not produce the optimal estimate $\hat{x}$. 
However, note that we do not need a very accurate estimate of this value but require the LQG 
control only to stabilize the entire control system. 
In fact with the above parameter choice, this purpose is fulfilled, and we end up with 
Fig.~\ref{LIGONDPAnoise_with_loss}; 
this shows that the proposed phase-cancellation filter can enhance the bandwidth in 
the high-frequency regime without sacrificing the peak-sensitivity.

\begin{table}[t]
\centering
\caption{Parameters used in Section VIII. 
The parameter $\gamma_\textrm{1loss}$, $\kappa_\textrm{4loss}$, and  $\kappa_\textrm{3loss}$ change in Fig.~\ref{controlled entire GW system}(a), (b), and (c), respectively. 
Note that $\kappa_1= 2 c / L_\textrm{arm}$ draws the connection between the GW interferometer and the phase-cancellation filter.
}\label{tab: parameters setup}
\begin{tabular}{|c|c|c|c|c|}
\hline
&Symbol & Definition & Value \\ \cline{1-4}
\multirow{10}*{\rotatebox[origin=c]{90}{GW interferometer}} &$M$ &mass of ETMs & 40 kg \\ \cline{2-4}
 &$L_\textrm{arm}$ & arm cavity length & 4 km \\ \cline{2-4}
 &$P_\textrm{arm}$ & arm cavity power & 800 kW \\ \cline{2-4}
 &\multirow{2}*{$\Omega_M$} & mechanical frequency  &\multirow{2}*{1 Hz} \\ 
 & & of ETMs &  \\ \cline{2-4}
  &$\lambda_\textrm{laser}$ &laser wavelength & 1064 nm \\ \cline{2-4}
 &$\omega_0$ & laser frequency & $=2\pi c / \lambda_\textrm{laser}$ \\ \cline{2-4}
 &$\Delta_d$ & effective detuning & $\approx-63$ Hz \\ \cline{2-4}
 & \multirow{2}*{$\gamma_\textrm{IFO}$} &effective coupling constant  & \multirow{2}*{$\approx1062$ Hz} \\ 
 & & between $d$ and $d_\textrm{in}$ & \\ \hline
\multirow{18}*{\rotatebox[origin=c]{90}{phase-cancellation filter}} &\multirow{2}*{$\lambda$} & coupling strength & \multirow{2}*{$3\times 10^6$ Hz} \\ 
 & & between $a_1$ and $a_2$ &  \\ \cline{2-4}
 &\multirow{3}*{$\gamma$} & coupling strength between  &\multirow{3}*{$=2.01\lambda$} \\
 &  &  the modes in NDPA and   & \\
 &  & the external itinerant fields  & \\ \cline{2-4}
 &\multirow{3}*{$\kappa_1$} & if $a_4$ was an itinerant field, & \multirow{3}*{$=2c/L_\textrm{arm}$} \\ \
 & &it represents the coupling   & \\ 
 & &strength between $a_3$ and $a_4$  &   \\ \cline{2-4} 
 &\multirow{2}*{$L_4$} & round trip cavity length of  & \multirow{2}*{0.5 m} \\ 
 & &  the cavity with the mode $a_4$  & \\ \cline{2-4}
 & \multirow{2}*{$g_{24}$} & coupling strength  &\multirow{2}*{$=\sqrt{c\gamma/L_4}$} \\ 
 & &  between $a_2$ and $a_4$  & \\ \cline{2-4}
 & \multirow{2}*{$g_{34}$} & coupling strength  &\multirow{2}*{$=\sqrt{c\kappa_1/L_4}$} \\ 
 & &  between $a_3$ and $a_4$  & \\ \cline{2-4}
 &\multirow{2}*{$\gamma_\textrm{1loss}$} & loss magnitude of  &\multirow{2}*{1 MHz} \\ 
 & &  the modes in NDPA & \\ \cline{2-4}
 &$\kappa_\textrm{3loss}$ & loss magnitude of $a_3$  & 100 Hz \\ \cline{2-4}
 &$\kappa_\textrm{4loss}$ & loss magnitude of $a_4$ & 600 kHz \\ \hline
\multirow{4}*{\rotatebox[origin=c]{90}{ Stabilizer}} &$Q$ &\multirow{2}*{regulator weights} & $I$ \\ \cline{2-2} \cline{4-4}
 &$R$ & & 0.01 \\ \cline{2-4}
 &\multirow{2}*{$V$} & \multirow{2}*{covariance matrix} & diag\{$10^{-22}$,\; 1/2, \\ 
 & & &\quad \quad \; $\cdots$,\; 1/2\} \\
\hline
\end{tabular}
\end{table}


\begin{figure}[t]

\begin{minipage}{\hsize}
\centerline{\hspace{-5.5em}\normalsize(a)}
\includegraphics[width=\columnwidth]{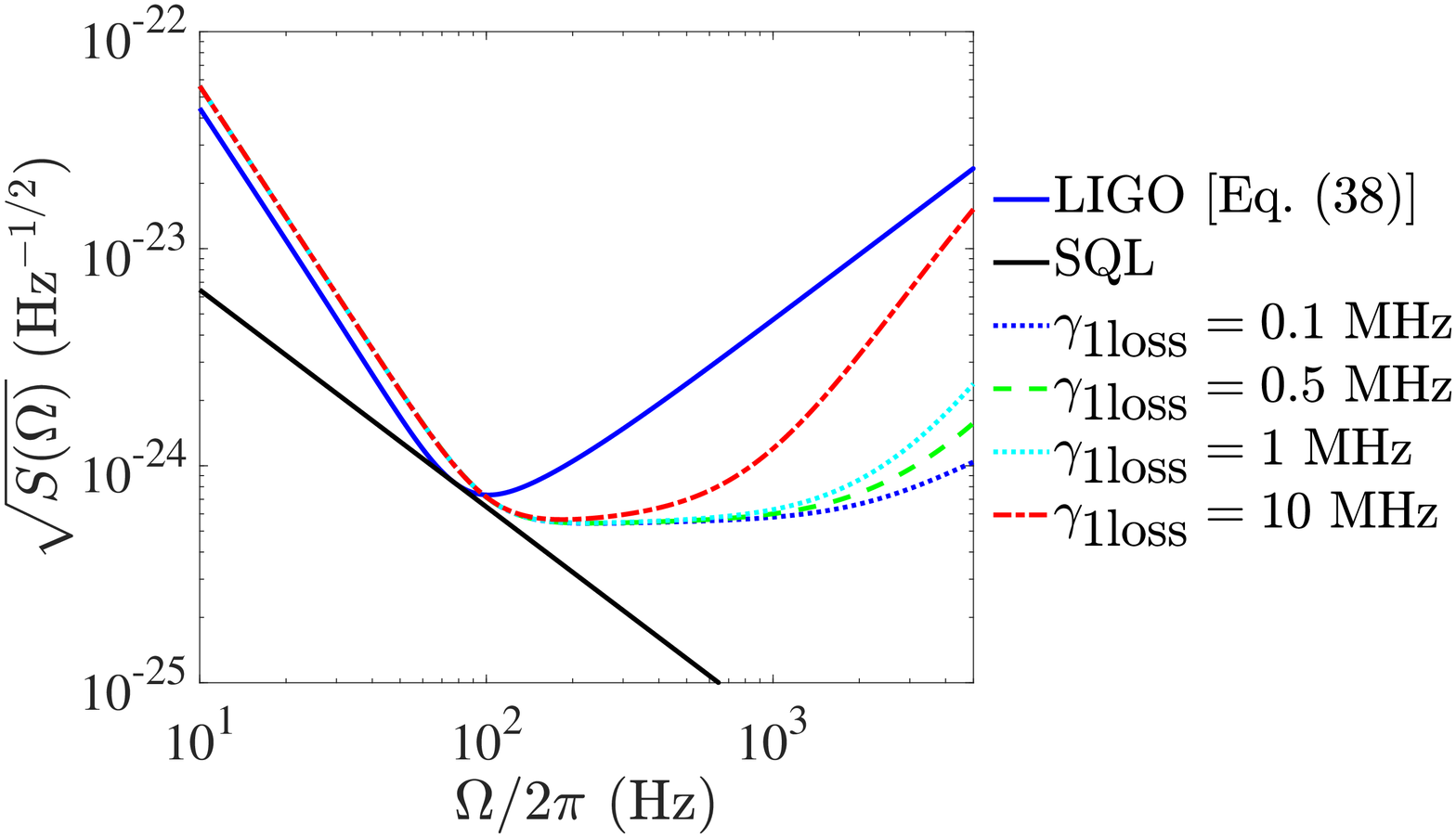}
\end{minipage}\\
\hspace{-0.5em}
\begin{minipage}{\hsize}
\vspace{1em}
\centerline{\hspace{-5.5em}\normalsize(b)}
\includegraphics[width=\columnwidth]{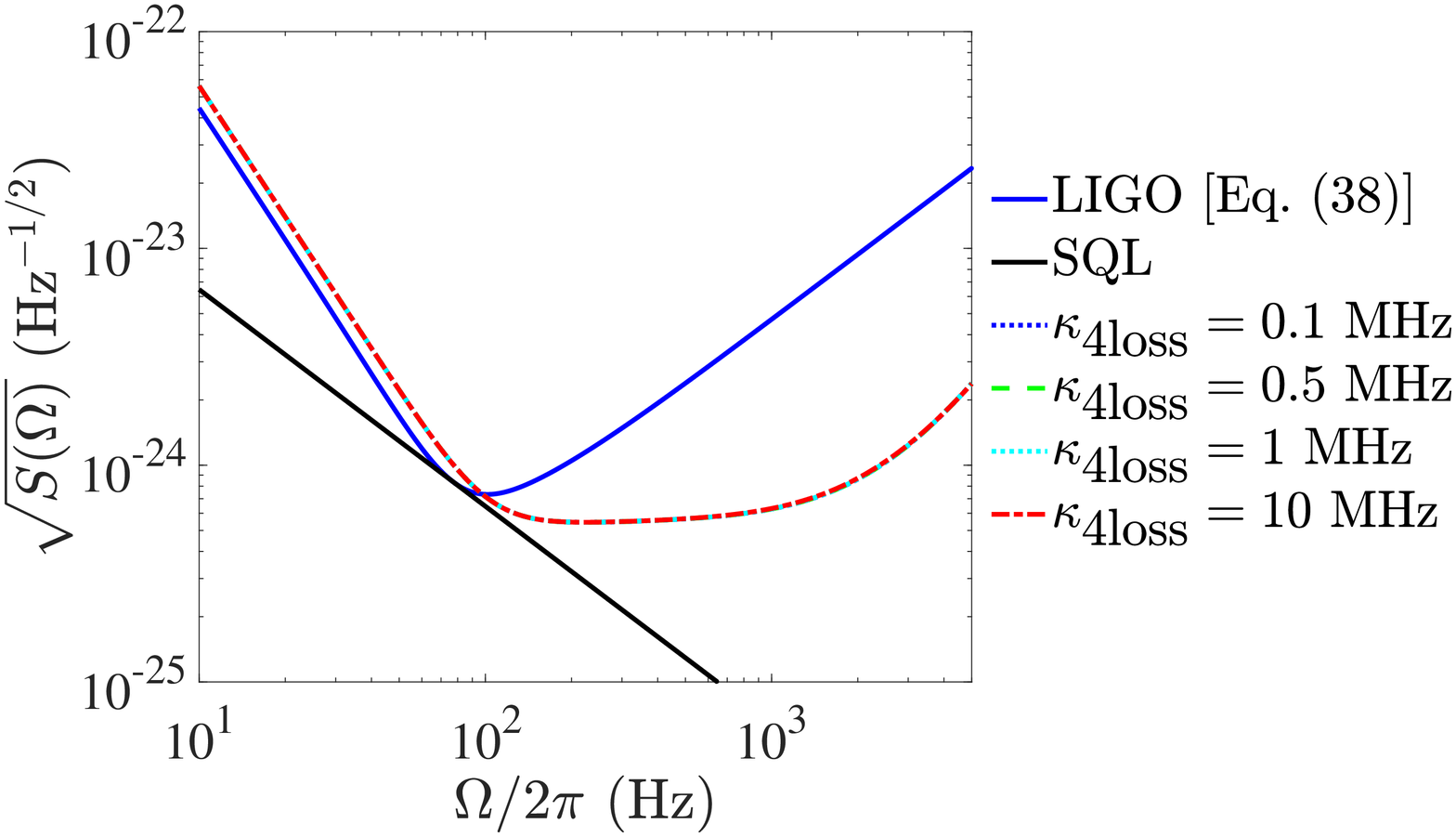}
\end{minipage}\\
\begin{minipage}{\hsize}
\vspace{1em}
\centerline{\hspace{-5.5em}\normalsize(c)}
\includegraphics[width=\columnwidth]{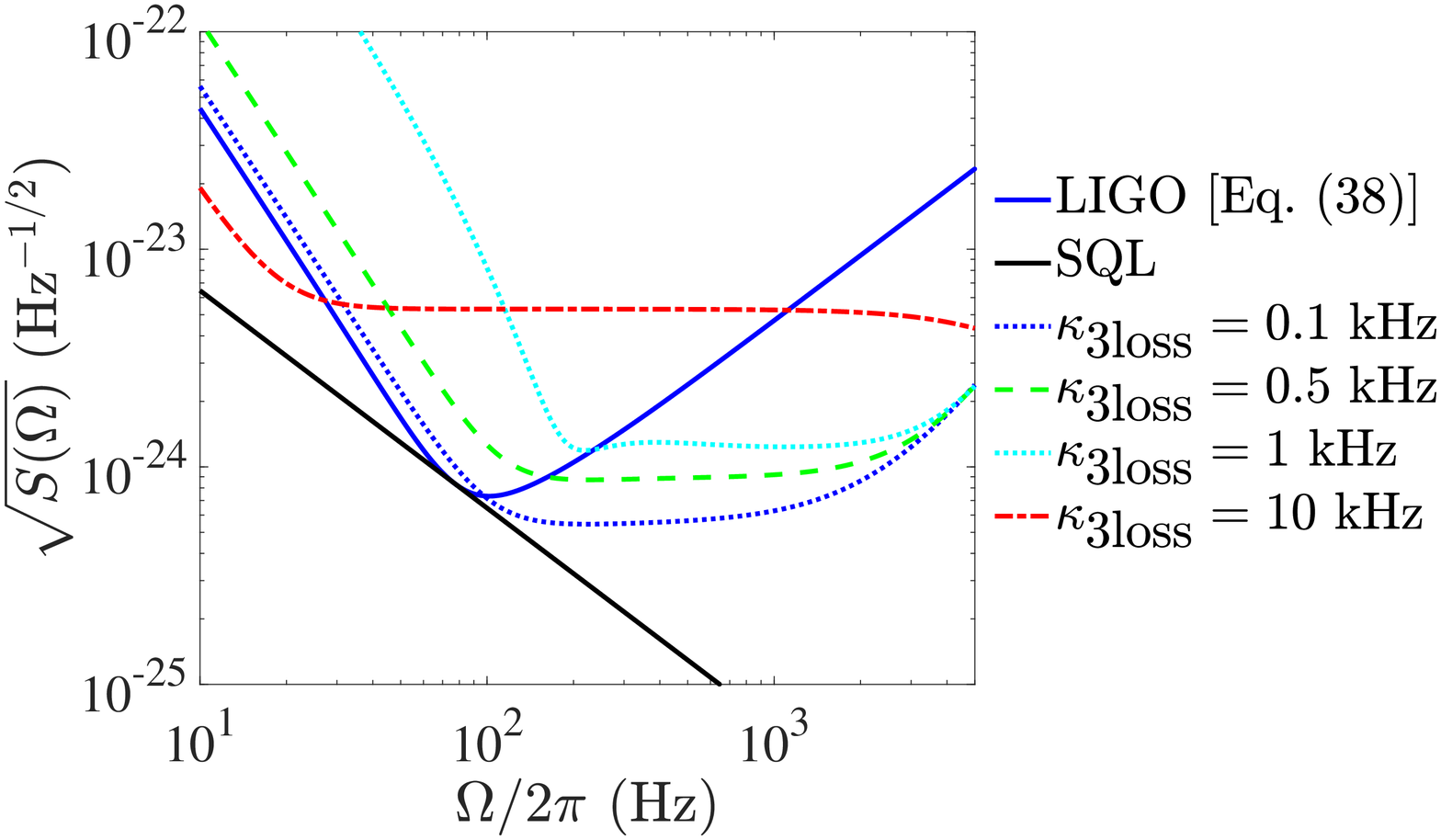}
\end{minipage}

\caption{
Quantum noise of the controlled gravitational-wave detector with several 
optical loss magnitudes of 
(a) the NDPA $\kappa_\textrm{1loss}$, (b) the loop cavity $\kappa_\textrm{4loss}$, 
and (c) the control cavity $\kappa_\textrm{3loss}$.}
\label{controlled entire GW system}
\end{figure}

We conclude this section with discussion on the possible advantages and 
disadvantages of the proposed filter. 
Figures~\ref{controlled entire GW system}(a) and (b) show the quantum noise 
of the entire controlled system with several optical loss magnitudes in 
(a) NDPA and (b) the loop cavity. 
Recall that the loss magnitude of the cavity modes are represented as  
$\gamma_\textrm{1loss}=cT_\textrm{1loss}/L_1$ and 
$\kappa_\textrm{4loss}=cT_\textrm{4loss}/L_4$, 
where ($T_\textrm{1loss}$, $T_\textrm{4loss}$) and ($L_1$, $L_4$) are the 
optical loss ratios and the round trip cavity lengths of the corresponding 
cavity modes, respectively. 
In the figures, the cavity lengths are fixed to $L_1=1.5$ m and $L_4=0.5$ m, 
and we change optical loss ratios to plot the noise spectral densities with 
several loss magnitudes $\gamma_\textrm{1loss}$ or $\kappa_\textrm{4loss}$. 
The other parameters are the same as those used in 
Fig.~\ref{LIGONDPAnoise_with_loss}. 
Importantly, the figures show that the sensitivity is not largely affected 
by the optical losses both in NDPA ($\gamma_\textrm{1loss}$) and the loop 
cavity ($\kappa_\textrm{4loss}$). 
In particular, the loss in the loop cavity has almost no effect on the 
sensitivity, as expected from the fact that the feedback amplification scheme 
is in general robust against the imperfection in the feedback loop 
\cite{Yamamoto 2016}. 
As for the loss in NDPA, there is certainly some impact on the sensitivity 
in the high frequency regime, but this can be reduced by making the length 
of NDPA longer. 

On the other hand, the parameter $\kappa_\textrm{3loss}$, i.e., the optical 
loss magnitude in the control cavity with mode $a_3$, has a large impact on 
the sensitivity, as indicated in Fig.~\ref{controlled entire GW system}(c); 
note that the parameters other than $\kappa_\textrm{3loss}$ are the same 
as those used in Fig.~\ref{LIGONDPAnoise_with_loss}. 
Figure~\ref{controlled entire GW system}(c) tells why $\kappa_\textrm{3loss}$ 
is chosen to be much smaller than $\gamma_\textrm{1loss}$ and $\kappa_\textrm{4loss}$ in Fig.~\ref{LIGONDPAnoise_with_loss}. 
To achieve such a small loss, the optical path length of the control cavity 
should be long; 
from $\kappa_\textrm{3loss}=cT_\textrm{3loss}/L_3$ with $T_\textrm{3loss}$ 
the loss ratio and $L_3$ the round trip length of the control cavity, if 
$\kappa_\textrm{3loss}=100$~Hz is required, we need, e.g., 
$T_\textrm{3loss}=0.01$\% and $L_3 = 300$~m. 
That is, although the proposed phase-cancellation filter based on the feedback amplification 
method can be constructed in all-optics way in contrast to the opto-mechanical proposal 
\cite{Miao 2015}, a very careful fabrication for the control cavity is required. 
In fact, to experimentally implement the proposed all-optics phase-cancellation filter requires 
a number of phase locks around the devices \cite{Tezak 2013, ida 2012}; also note that the large power 
level of the laser injected to the interferometer is required.


\section{Conclusion}
\label{sec:Conclusion}

In this paper, we have shown that a variety of quantum functionalities are generated 
under the concept of feedback amplification. 
We hope that, combined with the several established quantum information methods 
such as entanglement generation \cite{Girvin 15} and analogue information processing 
\cite{Devoret 10b}, those basic functionalities may be effectively applied to enhance 
the performance of existing quantum technological devices and moreover to create an
useful quantum mechanical machine.

\begin{acknowledgments}
This work was supported in part by JST PRESTO Grant No. JPMJPR166A. 
N.Y. acknowledges helpful discussions with H. Yonezawa, E. Huntington, M. Woolley, 
I. Petersen, M. James, and V. Ugrinovskii. 
\end{acknowledgments}


\appendix

\section{Proof of Eq.~\eqref{EQ:NON-RECIPROCAL AMP TF}}
\label{sec:Proof of Eq non-reciprocal amp tf}
The proof is composed of the following four steps. 
\begin{description}
\item[Step 1:] 
Derive the relation between $(b_1,b_2,b_3)$ and $(\tilde{b}_1,\tilde{b}_2,\tilde{b}_3)$ by using 
Eqs.~\eqref{eq: G in non-reciprocal amp}, \eqref{eq: Gbar in non-reciprocal amp}, and 
\eqref{eq: K in non-reciprocal amp}. 
More precisely, we find the transfer function $H(s)$ that satisfies the following relation:
\begin{align}
          \left[ \begin{array}{c}
                \tilde{b}_1(s) \\ 
                \tilde{b}_2^\dagger(s^\ast) \\
                \tilde{b}_3(s) \\ 
            \end{array} \right]
        &= H(s)
            \left[ \begin{array}{c}
                b_1^\dag(s^\ast) \\ 
                b_2(s) \\
                b_3(s) \\ 
            \end{array} \right],  \label{eq: H relation}\\
         H(s)=&
            \left[\begin{array}{ccc}
               H_{11}(s) &H_{12}(s) &H_{13}(s)  \\
               H_{21}(s) &H_{22}(s) & H_{23}(s) \\
               H_{31}(s) &H_{32}(s) & H_{33}(s)
            \end{array}\right].  \label{eq: H}
\end{align}
\item[Step 2:] Prove that, in the high gain limit regime, 
\begin{equation}
    H(s)
    \to
        \left[ \begin{array}{ccc}
            0 & -1 & 0 \\
            -1/K_{22}(s) &0   & -K_{21}(s)/K_{22}(s) \\
            -K_{12}(s)/K_{22}(s)& 0  & \det{[K(s)]}/K_{22}(s)\\
        \end{array} \right]. 
    \label{eq: H in high gain limit}
 \end{equation}
\item[Step 3:] 
Derive the relation between $(\tilde{b}_2, \tilde{b}_3,\tilde{b}_4)$ and $(b_1, b_3, b_4)$ using 
Eqs.~\eqref{eq: Kast in non-reciprocal amp}, \eqref{eq: H relation}, and \eqref{eq: H}. 
That is, we aim to have the expression of $G^\textrm{(fb)}(s)$ in terms of 
$\{H_{ij}(s)\}$ and $\{K_{ij}^\ast(s)\}$. 
\item[Step 4:] 
Substitute $\{H_{ij}(s)\}$ in Eq.~\eqref{eq: H in high gain limit} to $\{ G^\textrm{(fb)}_{ij}(s)\}$ obtained 
in Step 3, which leads to Eq.~\eqref{EQ:NON-RECIPROCAL AMP TF}. 
\end{description}
First, in Step 1, from Eqs.~\eqref{eq: G in non-reciprocal amp}, \eqref{eq: Gbar in non-reciprocal amp}, 
and \eqref{eq: K in non-reciprocal amp}, the entries in $H(s)$ are found to be 
\begin{align}
    H_{11}=&
        \frac{G_{12}-\overline{G}_{21}K_{22}\det{[G]}}{1-\overline{G}_{21}G_{21}K_{22}},~~
    H_{12}=
        \frac{G_{11}\overline{G}_{22}K_{22}}{1-\overline{G}_{21}G_{21}K_{22}},\notag\\
    H_{13}=&
        \frac{G_{11}K_{21}}{1-\overline{G}_{21}G_{21}K_{22}},~~
    H_{21}=
        \frac{\overline{G}_{11}G_{22}}{1-\overline{G}_{21}G_{21}K_{22}},\notag\\
    H_{22}=
        &\frac{\overline{G}_{12}+G_{21}K_{22}\det{[\overline{G}]}}{1-\overline{G}_{21}G_{21}K_{22}},~~
    H_{23}=
        \frac{\overline{G}_{11}G_{21}K_{21}}{1-\overline{G}_{21}G_{21}K_{22}},\notag\\
    H_{31}=&
        \frac{K_{12}\overline{G}_{21}G_{22}}{1-\overline{G}_{21}G_{21}K_{22}},~~
    H_{32}=
        \frac{K_{12}\overline{G}_{22}}{1-\overline{G}_{21}G_{21}K_{22}},\notag\\
    H_{33}=&
        \frac{K_{11}-\overline{G}_{21}G_{21}\det{[K]}}{1-\overline{G}_{21}G_{21}K_{22}},\notag
\end{align}
where we have omitted the Laplace index $s$ for simplicity. 
\\
\indent 
The proof of Step 2 is similar to that for deriving Eq.~\eqref{main result} in Sec.~\ref{sec:Feedback control}. 
That is, we take the ``quantum ideal op-amp assumption" as follows;
\begin{align}
     &\frac{\det{[G(s)]}}{G_{22}(s)} \to 0, ~~
     \frac{G_{12}(s)}{G_{22}(s)} \to 1, ~~
     \frac{G_{21}(s)}{G_{22}(s)} \to 1,\notag\\
     &\frac{\det{[\overline{G}(s)]}}{\overline{G}_{22}(s)} \to 0, ~~
     \frac{\overline{G}_{12}(s)}{\overline{G}_{22}(s)} \to 1, ~~
     \frac{\overline{G}_{21}(s)}{\overline{G}_{22}(s)} \to 1,
\notag
\end{align}
and $G_{11}(s)=G_{22}(s)$ and $\overline{G}_{11}(s)=\overline{G}_{22}(s)$ in the domain 
$s\in{\mathbb C}$ such that $|G_{11}(s)|\to\infty$ 
$(\Longleftrightarrow |G_{22}(s)|\to\infty)$ and $|\overline{G}_{11}(s)|\to\infty$ 
$(\Longleftrightarrow |\overline{G}_{22}(s)|\to\infty)$. 
Then in this high gain limit, the transfer functions are calculated as follows; 
\begin{align}
H_{11}=&
    \frac{G_{12}-\overline{G}_{21}K_{22}\det{[G]}}{1-\overline{G}_{21}G_{21}K_{22}}\notag\\
    =&
    \frac{(G_{12}/G_{22})/\overline{G}_{22}-(\overline{G}_{21}/\overline{G}_{22})K_{22}(\det{[G]}/G_{22})}{1/(G_{22}\overline{G}_{22})-(\overline{G}_{21}/\overline{G}_{22})(G_{21}/G_{22})K_{22}}\notag\\
    \to&0,\notag\\
H_{12}
    =&\frac{G_{11}\overline{G}_{22}K_{22}}{1-\overline{G}_{21}G_{21}K_{22}}\notag\\
    =&\frac{K_{22}}{1/(G_{22}\overline{G}_{22})-(\overline{G}_{21}/\overline{G}_{22})(G_{21}/G_{22})K_{22}}\to 1,\notag\\
H_{13}
    =&\frac{G_{11}K_{21}}{1-\overline{G}_{21}G_{21}K_{22}}\notag\\
    =&\frac{K_{21}/\overline{G}_{22}}{1/(G_{22}\overline{G}_{22})-(\overline{G}_{21}/\overline{G}_{22})(G_{21}/G_{22})K_{22}}\to0,\notag\\
H_{21}
    =&\frac{\overline{G}_{11}G_{22}}{1-\overline{G}_{21}G_{21}K_{22}}\notag\\
    =&\frac{1}{1/(G_{22}\overline{G}_{22})-(\overline{G}_{21}/\overline{G}_{22})(G_{21}/G_{22})K_{22}}\to -\frac{1}{K_{22}},\notag\\
H_{22}
    =&\frac{\overline{G}_{12}+G_{21}K_{22}\det{[\overline{G}]}}{1-\overline{G}_{21}G_{21}K_{22}}\notag\\
    =&\frac{(\overline{G}_{12}/\overline{G}_{22})/G_{22}+(G_{21}/G_{22})K_{22}(\det{[\overline{G}]}/\overline{G}_{22})}{1/(G_{22}\overline{G}_{22})-(\overline{G}_{21}/\overline{G}_{22})(G_{21}/G_{22})K_{22}}\notag\\
    \to&0,\notag\\
H_{23}
    =&\frac{\overline{G}_{11}G_{21}K_{21}}{1-\overline{G}_{21}G_{21}K_{22}}\notag\\
    =&\frac{(G_{21}/G_{22})K_{21}}{1/(G_{22}\overline{G}_{22})-(\overline{G}_{21}/\overline{G}_{22})(G_{21}/G_{22})K_{22}}\to -\frac{K_{21}}{K_{22}},\notag\\
H_{31}
    =&\frac{\overline{G}_{21}G_{22}K_{12}}{1-\overline{G}_{21}G_{21}K_{22}}\notag\\
     =&\frac{(\overline{G}_{21}/\overline{G}_{22})K_{12}}{1/(G_{22}\overline{G}_{22})-(\overline{G}_{21}/\overline{G}_{22})(G_{21}/G_{22})K_{22}}\to-\frac{K_{12}}{K_{22}},\notag\\
H_{32}
    =&\frac{\overline{G}_{22}K_{12}}{1-\overline{G}_{21}G_{21}K_{22}}\notag\\
     =&\frac{K_{12}/G_{22}}{1/(G_{22}\overline{G}_{22})-(\overline{G}_{21}/\overline{G}_{22})(G_{21}/G_{22})K_{22}}\to0,\notag\\
H_{33}
    =&\frac{K_{11}-\overline{G}_{21}G_{21}\det{[K]}}{1-\overline{G}_{21}G_{21}K_{22}}\notag\\
     =&\frac{K_{11}/(G_{22}\overline{G}_{22})-(\overline{G}_{21}/\overline{G}_{22})(G_{21}/G_{22})\det{[K]}}{1/(G_{22}\overline{G}_{22})-(\overline{G}_{21}/\overline{G}_{22})(G_{21}/G_{22})K_{22}}\notag\\
     \to& \frac{\det{[K]}}{K_{22}}.\notag
\end{align}
Now we have proved Eq.~\eqref{eq: H in high gain limit}. 
\\
\indent 
Step 3 can be completed by combining Eq.~\eqref{eq: Kast in non-reciprocal amp}, \eqref{eq: H relation}, 
and \eqref{eq: H}. 
The resulting expressions are found to be 
\begin{align}
G_{11}^\textrm{(fb)}
    = &
        \frac{
            H_{21}
            + (
                    H_{11}H_{22} - H_{12} H_{21}
              )
            K_{22}^\ast
            }{
                1 -H_{12} K_{22}^\ast
            },
    \label{eq: g11 h} \\
G_{12}^\textrm{(fb)}
    = &
        \frac{
            H_{23} + ( H_{13}H_{22}- H_{12} H_{23} )  K_{22}^\ast
        }{
            1 - H_{12} K_{22}^\ast
        },
     \\
G_{13}^\textrm{(fb)}
    = &
        \frac{
            H_{22}K_{12}^\ast
        }{
            1 - H_{12} K_{22}^\ast
        },
     \\
G_{21}^\textrm{(fb)}
    = &
        \frac{
            H_{31}
            +
            (
                H_{11}H_{32}
                -
                H_{12} H_{31}
            )
            K_{22}^\ast
        }{
            1 - H_{12} K_{22}^\ast
        },
     \\
G_{22}^\textrm{(fb)}
    = &
        \frac{
            H_{33}
            +
            (
                H_{13}H_{32}
                -
                H_{12} H_{33}
            )
            K_{22}^\ast
        }{
            1 - H_{12} K_{22}^\ast
        },
     \\
G_{23}^\textrm{(fb)}
    = &
        \frac{H_{32}K_{12}^\ast}{1 - H_{12} K_{22}^\ast},
     \\
G_{31}^\textrm{(fb)}
    = &
        \frac{H_{11} K_{21}^\ast}{1 - H_{12} K_{22}^\ast},
     \\
G_{32}^\textrm{(fb)}
    = &
        \frac{H_{13}K_{21}^\ast}{1 - H_{12} K_{22}^\ast},
     \\
G_{33}^\textrm{(fb)}
    = &
        \frac{
            K_{11}^\ast + H_{12}
            \big(
                K_{12}^\ast K_{21}^\ast - K_{11}^\ast K_{22}^\ast
            \big)
        }{
            1 - H_{12} K_{22}^\ast
        }.
    \label{eq: g33 h}
\end{align}
Step 4 is done by simply applying Eq.~\eqref{eq: H in high gain limit} to the above equations 
from Eq.~\eqref{eq: g11 h} to Eq.~\eqref{eq: g33 h}, which leads to 
Eq.~\eqref{EQ:NON-RECIPROCAL AMP TF}.


\begin{widetext}
\section{The matrix entries of $A$, $B_w$, $C$, $D$}
\label{sec:The matrix entries of ABCD}

\begin{equation}
A=\left[ 
\begin{array}{cccc cccc cccc}
0 &\Omega_M &0 &0 &0 &0 &0 &0 &0 &0 &0 &0  \\
0 &0 &\sqrt{2}G_M &0 &0 &0 &0 &0 &0 &0 &0 &0  \\
0 &0 &-\gamma_\textrm{IFO}/2 &\Delta &0 &g_\textrm{NI} &0 &0 &0 &0 &0 &0  \\
\sqrt{2}G_M &0 &-\Delta &-\gamma_\textrm{IFO}/2 &-g_\textrm{NI} &0 &0 &0 &0 &0 &0 &0  \\
0 &0 &0 &g_\textrm{NI} &-\gamma_\textrm{1loss}/2 &0 &\lambda &0 &0 &0 &0 &0  \\
0 &0 &-g_\textrm{NI} &0 &0 &-\gamma_\textrm{1loss}/2 &0 &-\lambda &0 &0 &0 &0  \\
0 &0 &0 &0 &\lambda &0 &0 &0 &0 &0 &0 &g_{24}  \\
0 &0 &0 &0 &0 &-\lambda &0 &0 &0 &0 &-g_{24} &0  \\
0 &0 &0 &0 &0 &0 &0 &0 &-\kappa_\textrm{3loss}/2 &0 &0 &g_{34}  \\
0 &0 &0 &0 &0 &0 &0 &0 &0 &-\kappa_\textrm{3loss}/2 &-g_{34} &0  \\
0 &0 &0 &0 &0 &0 &0 &g_{24} &0 &g_{34} &-\kappa_\textrm{4loss}/2 &0  \\
0 &0 &0 &0 &0 &0 &-g_{24} &0 &-g_{34} &0 &0 &-\kappa_\textrm{4loss}/2 
\end{array}
\right],\notag
\end{equation}

\begin{equation}
B_w=\left[
\begin{array}{ccccccccc}
0 &0 &0 &0 &0 &0 &0 &0 &0  \\
1/\sqrt{\hbar M\Omega_M} &0 &0 &0 &0 &0 &0 &0 &0  \\
0 &-\sqrt{\gamma_\textrm{IFO}} &0 &0 &0 &0 &0 &0 &0  \\
0 &0 &-\sqrt{\gamma_\textrm{IFO}} &0 &0 &0 &0 &0 &0  \\
0 &0 &0 &-\sqrt{\gamma_\textrm{1loss}} &0 &0 &0 &0 &0  \\
0 &0 &0 &0 &-\sqrt{\gamma_\textrm{1loss}} &0 &0 &0 &0  \\
0 &0 &0 &0 &0 &0 &0 &0 &0  \\
0 &0 &0 &0 &0 &0 &0 &0 &0  \\
0 &0 &0 &0 &0 &-\sqrt{\kappa_\textrm{3loss}} &0 &0 &0  \\
0 &0 &0 &0 &0 &0 &-\sqrt{\kappa_\textrm{3loss}} &0 &0  \\
0 &0 &0 &0 &0 &0 &0 &-\sqrt{\kappa_\textrm{4loss}} &0  \\
0 &0 &0 &0 &0 &0 &0 &0 &-\sqrt{\kappa_\textrm{4loss}} 
\end{array}
\right],\notag
\end{equation}
\begin{equation}
C=\left[
\begin{array}{cccccccccccc}
0 &0 &\sqrt{\gamma_\textrm{IFO}} &0 &0 &0 &0 &0 &0 &0 &0 &0  \\
0 &0 &0 &\sqrt{\gamma_\textrm{IFO}} &0 &0 &0 &0 &0 &0 &0 &0 
\end{array}
\right],\quad
D=\left[
\begin{array}{ccccccccc}
0 &1 &0 &0 &0 &0 &0 &0 &0  \\
0 &0 &1 &0 &0 &0 &0 &0 &0 
\end{array}
\right]\notag,
\end{equation}
where $G_M=G_\textrm{arm}\sqrt{\hbar/(M\Omega_M)}$.
\end{widetext}


\end{document}